\documentclass[sigconf,authorversion]{acmart}

\AtBeginDocument{%
  \providecommand\BibTeX{{%
    \normalfont B\kern-0.5em{\scshape i\kern-0.25em b}\kern-0.8em\TeX}}}

\usepackage{tikz}
\usepackage{xcolor}
\usepackage{multirow}
\usepackage{enumitem}
\usepackage{balance}
\usepackage{subcaption}
\usepackage{marvosym}

\copyrightyear{2024}
\acmYear{2024}
\setcopyright{acmlicensed}\acmConference[KDD '24]{Proceedings of the 30th ACM SIGKDD Conference on Knowledge Discovery and Data Mining}{August 25--29, 2024}{Barcelona, Spain}
\acmBooktitle{Proceedings of the 30th ACM SIGKDD Conference on Knowledge Discovery and Data Mining (KDD '24), August 25--29, 2024, Barcelona, Spain}
\acmDOI{10.1145/3637528.3671473}
\acmISBN{979-8-4007-0490-1/24/08}

\begin{document}

\title[Multimodal Pretraining, Adaptation, and Generation for Recommendation: A Survey]{Multimodal Pretraining, Adaptation, and Generation \\for Recommendation: A Survey}

\author{Qijiong Liu}
\authornote{\hspace{0.5ex}Equal contribution. Correspondence to: Jieming Zhu.}
\authornotemark[2]
\affiliation{%
  \institution{The HK PolyU}
  \city{Hong Kong}
  \country{China}
}
\email{liu@qijiong.work}

\author{Jieming Zhu}
\authornotemark[1]
\affiliation{%
  \institution{Huawei Noah's Ark Lab}
  \city{Shenzhen}
  \country{China}
  \country{}
}
\email{jiemingzhu@ieee.org}

\author{Yanting Yang}
\authornote{\hspace{0.5ex}The work was done when the authors were visiting at Huawei Noah's Ark Lab.}
\affiliation{%
  \institution{Zhejiang Unversity}
  \city{Hangzhou}
  \country{China}
  }
\email{yantingyang@zju.edu.cn}

\author{Quanyu Dai}
\affiliation{%
  \institution{Huawei Noah's Ark Lab}
  \city{Shenzhen}
  \country{China}
}
\email{daiquanyu@huawei.com}

\author{Zhaocheng Du}
\affiliation{%
  \institution{Huawei Noah's Ark Lab}
  \city{Shenzhen}
  \country{China}
}
\email{zhaochengdu@huawei.com}

\author{Xiao-Ming Wu}
\affiliation{%
  \institution{The HK PolyU}
  \city{Hong Kong}
  \country{China}
}
\email{xiao-ming.wu@polyu.edu.hk}

\author{Zhou Zhao}
\affiliation{%
  \institution{Zhejiang Unversity}
  \city{Hangzhou}
  \country{China}
  }
\email{zhaozhou@zju.edu.cn}

\author{Rui Zhang}
\affiliation{%
  \institution{Huazhong University of Science and Technology, China}
  \country{}
}

\email{rayteam@yeah.net}

\author{Zhenhua Dong}
\affiliation{%
  \institution{Huawei Noah's Ark Lab}
  \city{Shenzhen}
  \country{China}
}
\email{dongzhenhua@huawei.com}

\renewcommand{\shortauthors}{Qijiong Liu et al.}

\begin{abstract}
Personalized recommendation serves as a ubiquitous channel for users to discover information tailored to their interests. However, traditional recommendation models primarily rely on unique IDs and categorical features for user-item matching, potentially overlooking the nuanced essence of raw item contents across multiple modalities such as text, image, audio, and video. This underutilization of multimodal data poses a limitation to recommender systems, especially in multimedia services like news, music, and short-video platforms. The recent advancements in large multimodal models offer new opportunities and challenges in developing content-aware recommender systems. This survey seeks to provide a comprehensive exploration of the latest advancements and future trajectories in multimodal pretraining, adaptation, and generation techniques, as well as their applications in enhancing recommender systems. Furthermore, we discuss current open challenges and opportunities for future research in this dynamic domain. We believe that this survey, alongside the curated resources\footnote{\hspace{0.5ex}Github repository: \url{https://mmrec.github.io/survey}}, will provide valuable insights to inspire further advancements in this evolving landscape. 
\end{abstract}


\maketitle

\section{Introduction}
Recommender systems have been widely employed in various online applications, including e-commerce websites, advertising systems, streaming services, and social media platforms, to deliver personalized recommendations to users. Their primary goal is to enhance user experience, boost user engagement, and facilitate the discovery of items tailored to individual interests. However, traditional recommendation models primarily rely on unique IDs (e.g., user/item IDs) and categorical features (e.g., tags) for user-item matching~\cite{BARS}, potentially overlooking the nuanced essence of raw item contents across multiple modalities such as text, image, audio, and video~\cite{YuanYSLFYPN23}. This underutilization of multimodal data poses a limitation to recommender systems, especially in multimedia services like news, music, and short-video platforms~\cite{DeldjooSCP20}. 

To tackle this limitation, researchers have extensively investigated multimodal recommendation techniques for over a decade, resulting in a large body of research work that explores the integration of multimodal item features into recommendation models. For a comprehensive review, interested readers can refer to recent surveys~\cite{MMSurvey,MMSurvey2,DeldjooSCP20,MMSurvey3}. These surveys primarily delve into techniques such as multimodal feature extraction~\cite{MMSurvey3}, feature representation~\cite{DeldjooSCP20}, feature interaction~\cite{MMSurvey2}, feature alignment~\cite{DeldjooSCP20}, feature enhancement~\cite{MMSurvey2}, and multimodal fusion~\cite{MMSurvey} for recommendation models. However, the majority of these approaches rely on extracted multimodal feature embeddings, leaving other aspects of multimodal pretraining and generation relatively unexplored. Nowadays, pretrained large models have gained significant popularity in the domains of natural language processing (NLP), computer vision (CV), and multimodal systems (MM). The emergence of language models like the GPT~\cite{gpt-3} and Llama~\cite{llama-2} series has ushered in a new era of capabilities for understanding and generating language, while the CV field has witnessed breakthroughs with models such as ViT~\cite{vit} and DINOv2~\cite{DINOv2}. Leveraging these successes in unimodal domains, the multimodal community has concentrated on on aligning content across different modalities, such as CLIP~\cite{clip}, CLAP~\cite{clap} and BLIP-2~\cite{blip-2}. Notably, the recent introduction of groundbreaking technologies, such as ChatGPT~\cite{chatgpt}, SD~\cite{SD}, and Sora~\cite{Sora}, has further advanced the generation capabilities of pretrained large models to unprecedented levels. These recent advancements in pretrained large multimodal models offer new opportunities and challenges in developing content-aware recommender systems. 

\begin{figure*}[htbp]
    \centering
    \resizebox{.95\linewidth}{!}{
    \begin{subfigure}[b]{0.32\textwidth}
        \includegraphics[width=\textwidth]{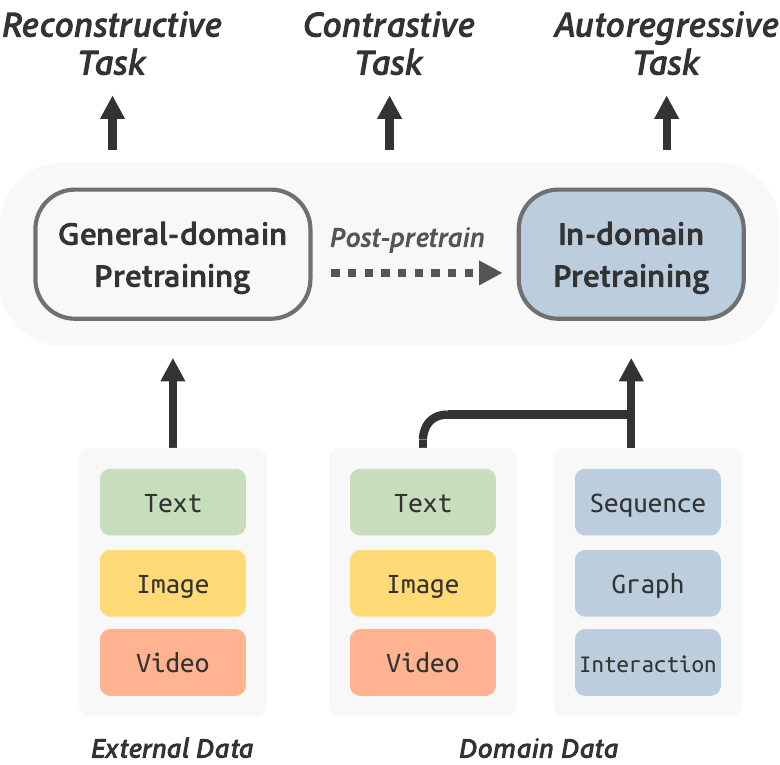}
        \caption{Multimodal Pretraining.}
        \label{fig:pretraining}
    \end{subfigure}
    \hspace{0.05\linewidth} 
    \begin{subfigure}[b]{0.32\textwidth}
        \includegraphics[width=\textwidth]{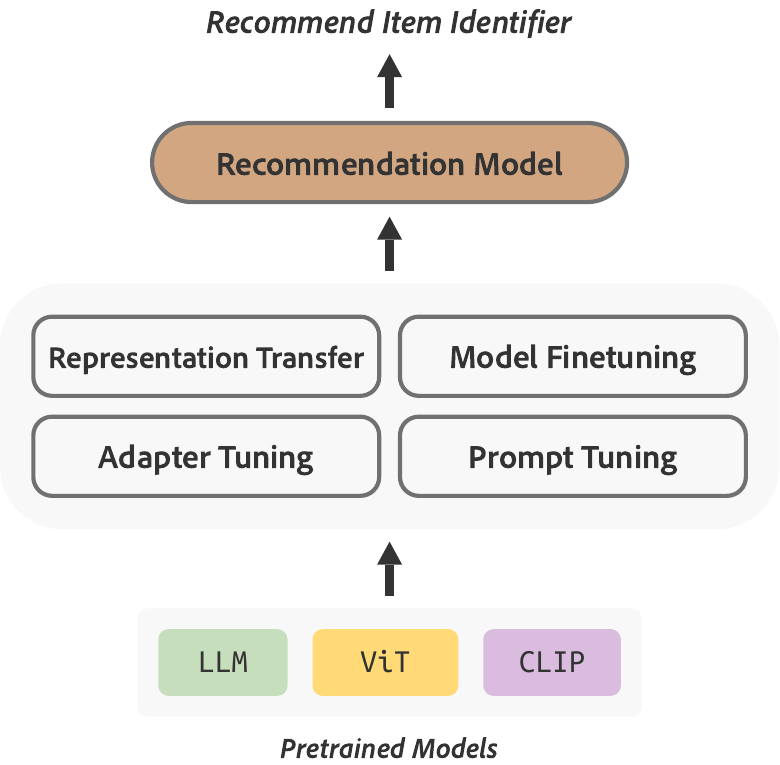}
        \caption{Multimodal Adaptation.}
        \label{fig:adaptation}
    \end{subfigure}
    \hspace{0.05\linewidth} 
    \begin{subfigure}[b]{0.32\textwidth}
        \includegraphics[width=\textwidth]{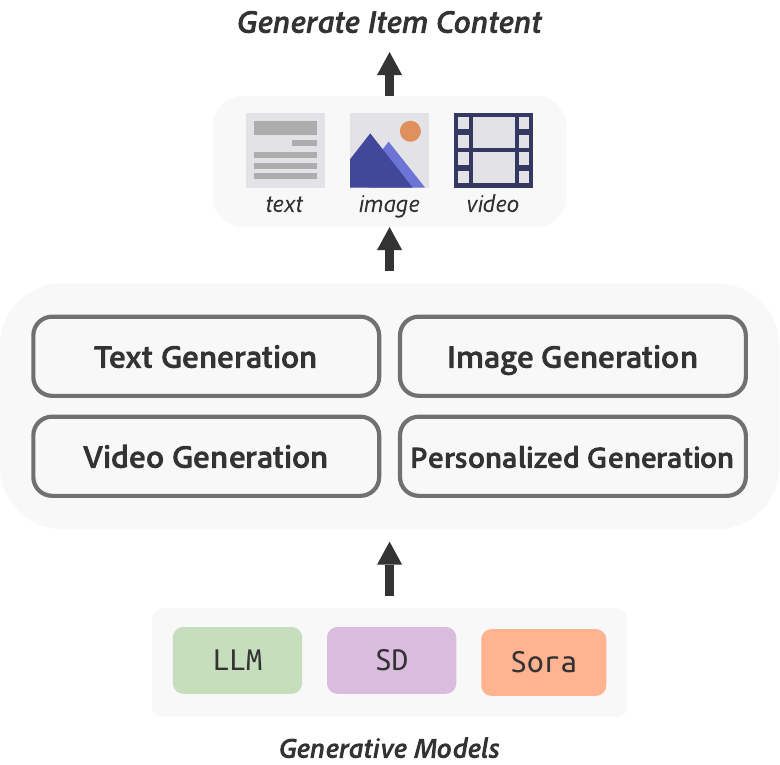}
        \caption{Multimodal Generation.}
        \label{fig:generation}
    \end{subfigure}
    }
    \caption{An overview of multimodal pretraining, adaptation, and generation tasks for recommendation.}
    \label{fig:overview}
\end{figure*}


In this survey, our objective is to provide a comprehensive overview of multimodal recommendation techniques from a new perspective, focusing on leveraging pretrained multimodal models. We explore the latest advancements and future trajectories in multimodal pretraining, adaptation, and generation techniques, along with their applications to recommender systems. Different from prior works, our survey capitalizes on recent progress in multimodal language models~\cite{blip-2,gpt-4}, prompt and adapter tuning~\cite{li2021prefix,zhou2022learning}, and generation techniques such as stable diffusion~\cite{SD}. Additionally, we delve into the most recent practical developments and remaining open challenges in applying pretrained multimodal models for recommendation tasks.

More specifically, Section \ref{sec:pretraining} introduces the task of multimodal pretraining for recommendation, emphasizing methods to enhance in-domain multimodal pretraining using domain-specific data. Section \ref{sec:adaptation} examines  multimodal adaptation for recommendation, elucidating  how pretrained multimodal models can be adapted to downstream recommendation tasks through techniques such as representation transfer, model finetuning, adapter tuning, and prompt tuning. Section \ref{sec:generation} delves into the emerging topic of multimodal generation for recommendation, with a focus on the application of AI-generated content (AIGC) techniques in recommendation contexts. In Section \ref{sec:application}, we outline a range of common applications that necessitate multimodal recommendation, followed by a discussion of open challenges and opportunities for future research in Section \ref{sec:challenge}. Finally, we conclude the survey in Section \ref{sec:conclusion}. We hope that this survey, along with the curated resources, will inspire further research efforts to advance this evolving landscape. 

\section{Multimodal Pretraining for Recommendation}\label{sec:pretraining}

In contrast to supervised learning directly on domain-specific data, self-supervised pretraining learns from a large-scale unlabeled corpus and then adapts the pretrained model to downstream tasks. This approach allows for the acquisition of rich external knowledge in pretraining data, thus leading to the widespread recognition of its effectiveness.
In this section, we will first provide a review of major pretraining paradigms and then introduce how they are utilized in the recommendation domain. Figure \ref{fig:pretraining} presents an overview of multimodal pretraining techniques.

\subsection{Self-supervised Pretraining Paradigms}

We broadly categorize self-supervised pretraining paradigms into three types according to their pretraining tasks. 

\textbf{Reconstructive Paradigm}. This pretraining paradigm aims to teach models to reconstruct raw inputs within the information bottleneck framework. Examples include \textit{mask prediction methods} for partial reconstruction and \textit{autoencoder methods} for complete reconstruction. 
Mask prediction methods were initially introduced in BERT~\cite{bert}, where input tokens are randomly masked, prompting the model to learn to predict them based on the surrounding context. In contrast, autoencoder methods (e.g., AE~\cite{ae}, VAE~\cite{vae}) encode input data into a concise latent space and subsequently learn to fully recover the input from this latent representation. These methods have found extensive use in self-supervised pretraining across various domains such as text~\cite{bart}, vision~\cite{vit, vq-vae, mae}, audio~\cite{musicbert}, and multimodal data~\cite{SD, dalle-2}. Following their success, researchers have applied the reconstructive pretraining paradigm to recommendation tasks. For instance, methods like mask item prediction in Bert4Rec~\cite{bert4rec}, mask token prediction in Recformer~\cite{recformer}, autoencoder-based item tokenization~\cite{tiger, uist}, and masked node feature reconstruction in PMGT~\cite{pmgt} have emerged. Despite significant progress, relying solely on this reconstructive paradigm may not capture proximity information from user-item interactions effectively. Consequently, these methods are typically complemented with a contrastive learning paradigm in practice.

\textbf{Contrastive Paradigm}. This pretraining focuses on pairwise similarity, distinguishing between similar and dissimilar data samples by maximizing distances between negative pairs and minimizing them for positive pairs within a representation space. It has proven effective in enhancing the quality of representations across different domains. Examples such as SimCSE~\cite{SimCSE} for text, SimCLR~\cite{SimCLR} for images, CLMR~\cite{clmr} for music, and CLIP~\cite{clip} for multimodal representations highlight its versatility and applicability. Given its ability to capture pairwise similarities, this paradigm finds extensive use in aligning user-item preferences. Applications like MGCL~\cite{mgcl}, MMSSL~\cite{mmssl}, MMCPR~\cite{liu2022multi}, MSM4SR~\cite{msm4sr}, and MISSRec~\cite{missrec} exemplify the utilization of contrastive learning for enhancing multimodal pretraining in recommender systems.

\textbf{Autoregressive Paradigm}. This paradigm has recently achieved remarkable success, particularly with the rise of large language models (LLMs) such as the GPT family~\cite{gpt, gpt-2, gpt-3, gpt-4}. It generates sequence data token by token in an autoregressive manner, where each token is predicted based on previous observations. In other words, this approach operates in a unidirectional, left-to-right generation framework, which is different from the reconstructive paradigm that employs bidirectional context to predict masked tokens. It has also gained rapid adoption in the CV domain~\cite{DALLE} and multimodal domain~\cite{AnyGPT}. In the realm of recommender systems, user behavior sequences naturally lend themselves to sequential processing, fostering the development of numerous autoregressive sequential recommendation models such as SASRec~\cite{sasrec}. Recent studies, such as P5~\cite{p5} and VIP5~\cite{vip5}, have explored the integration of LLMs or pretrained multimodal models into recommendation tasks. Concurrently, generative recommendation, which frames recommendation as autoregressive sequence generation, has emerged as a burgeoning area of research~\cite{tiger, eager, DeepMP}.


\subsection{Content-aware Pretraining for Recommendation}

Content-aware recommender systems strive to incorporate the semantic content of items to improve recommendation accuracy. Consequently, numerous studies have explored content-enhanced pretraining methods for recommendation systems. In this section, we categorize existing research based on the modalities employed for pretraining and discuss their application both generally and within recommendation systems.

\textbf{Text-based Pretraining}. 
Texts are among the most prevalent forms of content in recommender systems, applied in contexts such as news recommendation and review-based recommendation. Within the domain of natural language processing (NLP), pretrained language models like BERT~\cite{bert} and T5~\cite{t5} have been developed to capture context-aware representations of text. These models typically follow a pretraining-finetuning paradigm tailored to specific tasks. Recently, large language models (LLMs) such as ChatGPT~\cite{chatgpt} and LLaMa~\cite{llama} have demonstrated significant capabilities in language-related tasks, leveraging techniques such as prompting and in-context learning. Building on their success, text-enhanced pretraining has gained traction in recommender systems. Notable examples include MINER~\cite{miner} for news recommendation, Recformer~\cite{recformer} for sequential recommendation, UniSRec~\cite{unisrec} for cross-domain recommendation, and P5~\cite{p5} for LLM-based interactive recommendation.


\textbf{Audio-based Pretraining}. 
Music recommendation represents a prominent scenario heavily reliant on audio modalities to capture content semantics. Analogous to the NLP domain, various pretraining techniques have been employed to enhance audio representations, including Wav2Vec~\cite{wav2vec}, MusicBert~\cite{musicbert}, MART~\cite{MART}, and MERT~\cite{MERT}. In the context of music recommendation, researchers explore leveraging these audio pretraining methods by utilizing user-item interactions as supervision signals to finetune music representations. For example, Chen et al.\cite{uae} propose learning user-audio embeddings from track data and user interests through contrastive learning techniques. Furthermore, Huang et al.\cite{huang2020large} integrate pairwise textual and audio features into a convolutional model to jointly learn content embeddings in a similarity metric. Interested readers can find additional examples in a comprehensive review paper~\cite{MusicSurvey}.

\textbf{Vision-based Pretraining}. Images and videos constitute the primary visual data in multimedia recommendation scenarios as ads, movies, and videos. In the CV domain, the evolution of vision-based pretraining has transitioned from CNN-based architectures like ResNet~\cite{resnet} to transformer architectures such as ViT~\cite{vit} and DINOv2~\cite{DINOv2}, enabling the extraction of versatile visual features. These pretrained models have significantly advanced vision-aware recommendation systems. Researchers like Liu et al.\cite{cscnn} and Chen et al.\cite{hccm} leverage pretrained CNN encoders with category priors to extract image features for industrial recommendation tasks, finetuning image encoders alongside recommendation models. Similarly, Wang et al.\cite{missrec} and Wei et al.\cite{DeepMP} utilize pretrained transformer encoders, such as ViT~\cite{clip}, to encode images and sequences of user behaviors. Vision foundation models continue to evolve rapidly, promising future applications in recommendation tasks. Looking ahead, there is a growing interest in exploring the use of newly pretrained models for recommendation tasks. For instance, leveraging pretrained video transformers like Video-LLaVA~\cite{VideoLLaVA} remains an unexplored area in building video recommendation models.


\textbf{Multimodal Pretraining}. In current literature, most studies tend to focus on modeling the primary modality of content, such as text for news recommendation~\cite{miner}, audio for music recommendation~\cite{uae}, and images for e-commerce recommendation~\cite{cscnn}. However, multimedia content inherently involves multiple modalities. For instance, news articles often include titles, descriptions, and accompanying images. Similarly, video recommendation involves handling visual frames, audio signals, and subtitles. 

Unlike single-modal techniques, multimodal models must capture both commonalities and complementary information across multimodal data sources through techniques like cross-modal alignment and fusion. In recent years, multimodal pretraining has seen rapid development, resulting in a plethora of pretrained models, including single-stream models (e.g., VL-BERT~\cite{vlbert}), dual-stream models (e.g., CLIP~\cite{clip}), and hybrid models (e.g., FLAVA~\cite{FLAVA} and CoCa~\cite{coca}). Recent research has also focused on achieving unified representations of multimodal data, exemplified by models like ImageBind~\cite{imagebind}, MetaTransformer~\cite{Meta-Transformer}, and UnifiedIO-2~\cite{Unified-IO2}. Another emerging trend involves integrating multimodal encoders with large language models, resulting in multimodal large language models such as BLIP-2~\cite{blip-2}, Flamingo~\cite{flamingo}, and Llava~\cite{llava}. These advancements offer promising opportunities for building modern multimodal recommender systems. 

Initial efforts in this direction include models like MISSRec~\cite{missrec}, Rec-GPT4V~\cite{rec-gpt4v}, MMSSL~\cite{mmssl}, MSM4SR~\cite{msm4sr}, and AlignRec~\cite{alignrec}. However, current research primarily focuses on utilizing off-the-shelf pretrained multimodal encoders and integrating them with sequence-based or graph-based pretraining techniques for handling recommendation data. There remains a gap in how to specifically pretrain domain-specific multimodal models tailored for recommendation tasks. Pioneering efforts have been made in the e-commerce domain, as evidenced by models such as M5Product~\cite{M5Product}, K3M~\cite{K3M}, ECLIP~\cite{ECLIP}, and CommerceMM~\cite{CommerceMM}. Future research is expected to further explore and advance in this direction.


\section{Multimodal Adaption for Recommendation}\label{sec:adaptation}

While most existing pretrained models are trained on general data corpora, adapting them for recommender systems requires strategic methods to fully utilize their learned knowledge. This section summarizes four major adaptation techniques: representation transfer, model finetuning, adapter tuning, and prompt tuning. Each technique provides a distinct approach to harnessing the benefits of a pretrained model. Figure \ref{fig:adaptation} presents an overview of multimodal adaptation techniques.

\subsection{Representation Transfer}

Representation transfer is one of the most commonly used adaptation techniques for transferring pretrained knowledge to recommendation models. Specifically, item representations are extracted from frozen pretrained models and used as additional features alongside ID embeddings. These representations provide supplementary general information to recommender systems, addressing the cold-start problem where new or infrequently interacted items may have inadequate ID embeddings derived from limited interactions~\cite{prec}. Approaches based on representation transfer have been extensively studied and proven effective across various domains, e.g., text-based recommendation~\cite{nrms}, vision-based recommendation~\cite{shang2023learning,itemsage}, multimodal recommendation~\cite{im-rec,kdsr,mtpr,lightgt,nova}. For multimodal scenarios specifically, significant efforts have been directed towards fusing representations from multiple modalities. This includes techniques such as early fusion~\cite{liu2019user, kdsr, pmgt, freedom}, intermediate fusion~\cite{im-rec, mtpr, mmrec}, and late fusion~\cite{mgat, grcn}. Another research focus involves aligning multimodal representations within user behavior spaces using methods like content-ID alignment~\cite{CLCRec,IDEmbedding}, item-to-item matching~\cite{ssd}, user sequence modeling~\cite{unisrec}, and graph neural networks~\cite{mmgcn}.

However, straightforward and efficient representation transfer may encounter a significant domain generalization gap due to misalignment between the semantic space and the behavior space, which may not consistently lead to performance improvements in practice. Model finetuning offers a direct solution to address this issue. Furthermore, as noted by KDSR~\cite{kdsr}, there is a risk of forgetting modality features in representations, leading them to resemble those trained without incorporating such features. One viable approach involves integrating explicit constraints~\cite{kdsr}, while another potential solution introduces semantic tokenization techniques~\cite{tiger, SemanticID, uist, CoST}, which quantize item content representations into discrete tokens.

\subsection{Model Finetuning}

Model finetuning refers to the process of further training a pretrained model on task-specific data. Its goal is to adapt the model parameters to effectively capture domain-specific nuances, thereby improving its performance on the specific downstream task. This pretraining-finetuning paradigm has proven successful in various practical applications. Specifically, finetuning can involve aligning the semantic space of pretrained models with the behavior space of recommendation models. Depending on the application of pretrained models, they can extract item representations~\cite{miner, xv2022visual}, user representations~\cite{sun2023universal}, or both~\cite{recformer, prec, missrec}. Moreover, based on the types of downstream tasks, current research can be classified into representation-based matching tasks~\cite{plm-nr, miner, missrec} and scoring-based ranking tasks~\cite{unbert, xv2022visual, cscnn}.

However, end-to-end finetuning of pretrained models with recommendation data faces challenges related to training efficiency. Recommendation tasks often require processing millions or even billions of samples daily. Fully finetuning a pretrained model substantially amplifies training overhead, which poses practical limitations in large-scale recommender systems, particularly given the scale of large language and multimodal models~\cite{llama, blip-2}. Moreover, finetuning with large volumes of data easily leads to the issue of catastrophic forgetting, where previously learned knowledge rapidly deteriorates during continual training.

\subsection{Adapter Tuning}


To reduce training overhead with pretrained large models, parameter-efficient finetuning (PEFT) methods have been developed. One prominent approach is through parameter-efficient adapters like LoRAs~\cite{lora}, which integrate compact, task-specific modules directly into pretrained models. This strategy effectively reduces the number of parameters needed for finetuning and facilitates rapid model adaptation. Widely recognized for its efficacy across various domains, PEFT techniques have gained significant traction in recommendation systems~\cite{once, unisrec, transrec, xi2023towards, vip5}. For example, the ONCE framework~\cite{once} leverages the pretrained Llama model~\cite{llama} with LoRAs as item encoders to enhance content-aware recommendation. Similarly, UniSRec~\cite{unisrec} employs an MOE-based adapter with the BERT model to improve semantic representations of items across diverse domains. In the realm of multimodal recommendation, TransRec~\cite{transrec} and VIP5~\cite{vip5} have introduced layerwise adapters. M3SRec utilizes modality-specific MOE adapters~\cite{M3SRec}, while EM3 employs multimodal fusion adapters~\cite{EM3} during finetuning. Nonetheless, the ongoing challenge lies in designing adapters that effectively balance both effectiveness and efficiency, which remains an active area of research.

\subsection{Prompt Tuning}

With the emergence of large language models, prompting has become a pivotal technique in harnessing their capabilities to generate desired outputs or perform specific tasks~\cite{PromptSuvey}. Instead of using handcrafted prompts, prompt tuning aims to learn task-adaptable prompts from task-specific data while keeping the model parameters frozen. As a result, prompt tuning can avoid catastrophic forgetting and enable fast adaptation with only prompt tokens as tunable parameters. Depending on whether prompts are optimized in a discrete token space, they can be further categorized into hard prompt tuning, such as AutoPrompt~\cite{shin2020autoprompt}, and soft prompt tuning, like Prefix-Tuning~\cite{li2021prefix}. Prompt tuning has been successful in visual learning~\cite{zhou2022learning} and multimodal learning~\cite{Crossmodalprompt, khattak2023maple}, enhancing model performance.

For multimodal recommendation tasks, prompt tuning has emerged as a novel technique to adapt pretrained models. Recent studies such as RecPrompt~\cite{liu2023recprompt}, ProLLM4Rec~\cite{xu2024prompting}, Prompt4NR~\cite{Prompt4NR}, and PBNR~\cite{li2023pbnr} employ prompting methods to customize large language models (LLMs) for news recommendation tasks. Additionally, DeepMP~\cite{DeepMP}, VIP5~\cite{vip5}, and PromptMM~\cite{wei2024promptmm} employ prompt tuning to integrate and adapt multimodal content knowledge to enhance recommendation. Despite recent advancements, this area of research remains underexplored. An interesting direction is the development of personalized multimodal prompting techniques to advance multimodal recommendation systems.

\section{Multimodal Generation for Recommendation}\label{sec:generation}

With recent advancements in generative models, AI-generated content (AIGC) has gained significant popularity across diverse applications. In this section, we explore potential research avenues for employing AIGC techniques within recommender systems. Figure \ref{fig:generation} presents an overview of multimodal generation techniques.

\subsection{Text Generation}
With the support of powerful large language models (LLMs), text generation has become a mature capability and is now being applied in various tasks within the recommendation domain~\cite{murakami2023natural}.

\begin{itemize}[leftmargin=*]
  \item {\textbf{Keyword Generation}: Keyword tagging plays a pivotal role in content understanding for ads targeting and recommendation. Previous techniques mostly rely on explicit keyword extraction from textual content, potentially missing important keywords absent from the text. Consequently, keyword generation techniques have been widely applied to enhance the keyword tagging process~\cite{ChatGPTKeyphrase,TagGPT}}. 
  \item \textbf{News Headline Generation}: The demand for personalized and engaging news content has fueled the exploration of news headline generation. Conventionally, headline generation is framed as a text summarization task, condensing input text or multimodal content into a title~\cite{NewsGen,DingSZTJ23}. However, typical news headlines may lack appeal or relevance to specific users, prompting the need for personalized approaches. Consequently, personalized headline generation has emerged as a compelling research topic, focusing on generating titles tailored to individual users' reading preferences and available news content~\cite{NHNet,LaMP}.
  \item \textbf{Marketing Copy Generation}: Marketing copy refers to the text used to promote a product and motivate consumers to purchase. It plays a vital role in capturing users' interest and enhancing engagement. Recent efforts have focused on automatic marketing copywriting based on LLMs~\cite{CAMERA,zhou2024gcof} 
  \item \textbf{Explanation Generation}: In interactive scenarios, the demand for explainable recommendations is growing significantly. This involves generating natural language explanations to justify the recommendation of items to individual users, thereby enhancing user understanding and trust in the system~\cite{li2023personalized,ZhaoFSSCXQ19}.
  \item \textbf{Dialogue Generation}: Dialogue generation~\cite{YangZEL22} is essential in conversational recommender systems, encompassing the generation of responses that describe recommended items~\cite{LLMCRS}. Moreover, it entails generating questions to guide users towards further rounds of conversation and interaction~\cite{WangTRCWA23}.
\end{itemize}

While these tasks benefit from powerful LLMs, two critical challenges persist in text generation for recommendation: 1) Controllable Generation: Industrial applications necessitate precise control over generated texts to ensure correctness of product descriptions, use unique selling propositions, or adhere to specific writing styles~\cite{ZhouJWCS23}. 2) Knowledge-Enhanced Generation: Existing LLMs often lack explicit awareness of domain-specific knowledge, such as product entities, categories, and selling points. Recent research has concentrated on integrating domain-specific knowledge bases to achieve more satisfactory results~\cite{KPLUG,TrisedyaQWZ22}.

\subsection{Image and Video Generation}
Text-to-image generation has achieved remarkable success with the prevalence of diffusion models (e.g., SD~\cite{SD}). In this section, we delve into their potential applications in e-commerce and advertising. Unlike natural image generation, generating product images and ad banners involves dealing with complex layouts, encompassing various elements such as products, logos, and textual descriptions. Consequently, unique challenges arise in designing a coherent layout and effectively integrating text with appropriate fonts and colors to create visually appealing posters.

Specifically, Inoue et al.~\cite{LayoutDM} propose LayoutDM, a model designed to effectively handle structured layout data and facilitate the discrete diffusion process. Hsu et al.~\cite{PosterLayout} enable content-aware layout generation (namely PosterLayout) by arranging predefined spatial elements on a given canvas. Lin et al. \cite{AutoPoster} develop AutoPoster, a highly automated and content-aware system for generating advertising posters. 
Concurrently, some studies explore text design for poster generation. For example, Gao et al.~\cite{TextPainter} introduce TextPainter, a novel multimodal approach that leverages contextual visual information and corresponding text semantics to generate text images. Tuo et al.~\cite{AnyText} propose a diffusion-based multilingual visual text generation and editing model, AnyText, which addresses how to render accurate and coherent text in the image.

More recently, video generation has made significant strides. Sora~\cite{Sora} emerges as a groundbreaking technology showcasing immense potential for generating advertising videos for products. In this context, Gong et al. \cite{AtomoVideo} introduce AtomoVideo, a high-fidelity image-to-video generation solution that effectively transforms product images into engaging promotional videos for advertising purposes. Additionally, Liu et al. \cite{LiuYDSYDGWRXCM20} have devised a system capable of automatically generating visual storylines from a given set of visual materials, producing compelling promotional videos tailored for e-commerce. Furthermore, Wang et al.~\cite{MagicVideo} have developed an integrated approach, merging text-to-image models, video motion generators, reference image embedding modules, and frame interpolation modules into an end-to-end video generation pipeline, which is valuable for micro-video recommendation platforms. We believe that this field is rapidly expanding, enabling the advancement of AIGC-based recommendation and advertising applications.

\subsection{Personalized Generation}
With the rise of AIGC, there is a notable shift towards personalized generation, aiming to enhance the customization and personalization of generated content. This trend holds particular significance in recommendation scenarios, where personalized content can better cater to users' interests. Pioneering work has been undertaken in various domains, including personalized news headline generation~\cite{PENS,PNG,LaMP,GUE}, personalized product description generation in e-commerce~\cite{DengLZDL22}, personalized answer generation~\cite{DengLZDL22}, personalized image generation with identity preservation~\cite{vit}, and personalized multimodal generation~\cite{PMG}. Integrating recommender systems with personalized generation techniques shows promise for developing next-generation recommender systems. 

\section{Applications}\label{sec:application}
In this section, we summarize some common application domains that require multimodal recommendation techniques. 

\begin{itemize}[leftmargin=*]
\item \textbf{E-commerce Recommendation}. E-commerce represents one of the most extensively studied application domains in recommender systems research, aimed at assisting users in discovering items they are likely to purchase. The abundance of multimodal data in e-commerce, including product titles, descriptions, images, and reviews, poses a challenge in integrating different modalities with user interaction data to enhance recommendation quality. To address this challenge, numerous research efforts have been undertaken. Notable examples include works by Alibaba~\cite{ImageMatters,xv2022visual,LiWTZOOZ20}, JD.com~\cite{XiaoDCJYDL22,cscnn}, and Pinterest~\cite{itemsage}.


\item \textbf{Advertisement Recommendation}. Online advertising serves as a primary revenue source for many web applications. Advertising creatives play a pivotal role in this ecosystem, spanning various formats such as images, titles, and videos. Aesthetic creatives have the potential to engage potential users and enhance the click-through rate (CTR) of products~\cite{AutomatedCreative}. There is also a pressing need to understand ad creatives better to effectively align advertisements with users' interests~\cite{YangDTTZQD19,ParallelRanking}.

\item \textbf{News Recommendation}. Personalized news recommendation is a crucial technique for assisting users in discovering news of interest. To enhance recommendation accuracy and diversity, recommender systems must comprehend news content and extract semantic information from a user's reading history. This often involves learning semantic representations of news titles, abstracts, body text, and cover images. Recent research has focused on modeling features from multiple modalities, as exemplified by MM-Rec~\cite{mmrec} and IMRec~\cite{im-rec}.



\item \textbf{Video Recommendation}.  
With the surge in popularity of micro-video platforms, video recommendation has garnered significant attention within the community. Videos encapsulate a multitude of modalities, including titles, thumbnail images, frames, audio tracks, transcripts, and more. Current research efforts have been concentrated on integrating and adapting multimodal information within micro-video recommendation models.~\cite{mmgcn,multimodalGC}. Notably, Ni et al.~\cite{MicroLens} have recently introduced a comprehensive micro-video recommendation dataset, enriched with abundant multimodal side information, to foster further research in this domain.



\item \textbf{Music Recommendation}. 
The realm of music streaming services represents another prominent domain that necessitates multimodal recommendation techniques. Within this sphere, a diverse array of multimodal data is involved, including music audio, scores, lyrics, tags, and reviews. Leveraging these various types of music data has proven effective in crafting more personalized recommendations aimed at engaging users; notable examples can be found in~\cite{huang2020large,uae}. Additionally, Shen et al.~\cite{shen2020enhancing} propose that incorporating multimodal information from users' social media can offer insights into their personalities, emotions, and mental well-being, thereby enhancing the accuracy of music recommendation.


\item \textbf{Fashion Recommendation}. With the visual and aesthetic nature of fashion products, fashion recommendation has emerged as a distinct vertical domain. Unlike traditional recommender systems, fashion recommendation not only suggests individual items but also outfits that complement multiple items. Multimodal understanding capabilities play a pivotal role in this area, including tasks such as localizing fashion items from images, identifying their attributes, and computing compatibility scores for multiple items~\cite{DeldjooNRMPBN24,Song2023MMFRecME}. Moreover, pioneering work~\cite{ZhuYZRCS0K23} has developed text-to-image diffusion models that allow users to virtually try on clothes. These techniques are expected to enhance the personalization of fashion recommendation and elevate user experience to the next level.





\item \textbf{LBS Recommendation}. Location-based services (LBS) have become ubiquitous, offering a wide range of services including taxi travel, food delivery, and restaurant recommendation. In these contexts, users can share their Points of Interest (POI) check-ins, photos, opinions, and comments, which encompass a rich array of multimodal spatio-temporal data. Integrating this multimodal information and understanding spatio-temporal correlations among locations enables more accurate modeling of user preferences. Notable examples can be found in~\cite{Qin2022NextPR,Liu2023MandariMT}.
\end{itemize}



\section{Challenges and Opportunities}\label{sec:challenge}

In this section, we discuss the 
persistent challenges and emerging opportunities for future research.
\begin{itemize}[leftmargin=*] 

\item \textbf{Multimodal Information Fusion}. Multimodal fusion has been extensively explored in research. Within recommender systems, current studies primarily concentrate on fusing and adapting multimodal feature embeddings of items to recommendation models~\cite{MMSurvey}. However, multimodal information for recommendation inherently adopts a hierarchical structure, ranging from user behavior sequences to individual items, each comprising multiple modalities and further subdivided into semantic tokens and objects. Additionally, the impact of information from diverse modalities and regions can vary significantly among different users. As a result, the challenge lies in effectively fusing multimodal information in a hierarchical and personalized manner to optimize recommendations.


\item \textbf{Multimodal Multi-domain  Recommendation.}
Multimodal information provides rich semantic insights into item content. Despite considerable research into multimodal recommendation and cross-domain recommendation, effectively leveraging multimodal information to bridge the information gap across domains remains an open challenge~\cite{sun2023universal}. For instance, recommending music based on a user's reading habits entails semantic alignment across modalities (audio vs. text) and domains (music vs. books).


\item \textbf{Multimodal Foundation Models for Recommendation.}
While large language models and large multimodal models have emerged as foundation models in the NLP and CV domains, there exists a compelling opportunity to extend this exploration into the recommendation domain. An ideal recommendation foundation model should demonstrate robust in-context learning capabilities while maintaining generalizability across diverse tasks and domains~\cite{FoundationSurvey}. Potential avenues for exploration include adapting existing multimodal LLMs for recommendation tasks (e.g., \cite{vip5}), or conducting the pretraining of a multimodal generative model from scratch using large-scale multimodal multi-domain recommendation data.

\item \textbf{AIGC for Recommendation.}
The integration of AIGC represents a notable advancement in recommender systems, offering an opportunity to significantly enhance user personalization, engagement, and overall experience. This encompasses personalized news headlines, tailored advertising creatives, and explanatory content across diverse recommendation contexts. This field is rapidly expanding, with the primary challenge lying in achieving a comprehensive understanding of both content and users, facilitating controllable generation, and ensuring accurate formatting to optimize the user experience. Additionally, it is imperative to address potential ethical and privacy concerns arising from the use of AIGC.


\item \textbf{Multimodal Recommendation Agent.}
LLM-based agents~\cite{voyager} have demonstrated exceptional proficiency in automating tasks through extensive knowledge and strong reasoning capabilities. The integration of these agents has introduced innovative prospects in the field of recommendation, particularly in conversational recommendation~\cite{fang2024multi}. This entails directly engaging users in the task completion process, thereby enhancing the user experience and the effectiveness of recommender systems. As a concrete example, integrating conversation and virtual try-on generation~\cite{ZhuYZRCS0K23} capabilities may present new opportunities for fashion recommendation.



\item \textbf{Efficiency of Training and Inference.}
Recommendation tasks typically have stringent latency requirements to meet real-time service demands. Therefore, ensuring training and inference efficiency becomes imperative when applying multimodal pretraining and generation techniques in practice. There is a high demand for the development of efficient strategies to leverage the capabilities of multimodal models. Pioneer efforts in this direction include speeding up training by merging item sets to avoid redundant encoding operations~\cite{SpeedyFeed} and enhancing inference speed~\cite{prec, hccm} through caching item and user representations.



\end{itemize}

\section{Conclusion}\label{sec:conclusion}

Multimodal recommendation, an immensely promising field, has garnered significant attention in recent years, fueled by advancements in both multimodal machine learning and the recommendation system community. The advent of large multimodal models has transformed the multimodal recommendation landscape, endowing it with enhanced capabilities for comprehension and content generation. This paper provides a systematical overview of the current multimodal recommendation framework, focusing on key aspects such as multimodal pretraining, adaptation, and generation. Additionally, we delve into its applications, challenges, and future prospects. Our aim is to offer this survey as a resourceful guide to aid subsequent research in the field.

\begin{acks}
We thank \href{https://xinzhou.me}{Dr. Xin Zhou} and \href{https://wuch15.github.io}{Dr. Chuhan Wu} for the discussion and contribution to the tutorial materials of multimodal pretraining and generation for recommendation presented at WWW 2024~\cite{mmrec_tutorial}.
\end{acks}

\balance
\bibliographystyle{ACM-Reference-Format}
\bibliography{MM4Rec}


\begin{thebibliography}{165}


\ifx \showCODEN    \undefined \def \showCODEN     #1{\unskip}     \fi
\ifx \showDOI      \undefined \def \showDOI       #1{#1}\fi
\ifx \showISBNx    \undefined \def \showISBNx     #1{\unskip}     \fi
\ifx \showISBNxiii \undefined \def \showISBNxiii  #1{\unskip}     \fi
\ifx \showISSN     \undefined \def \showISSN      #1{\unskip}     \fi
\ifx \showLCCN     \undefined \def \showLCCN      #1{\unskip}     \fi
\ifx \shownote     \undefined \def \shownote      #1{#1}          \fi
\ifx \showarticletitle \undefined \def \showarticletitle #1{#1}   \fi
\ifx \showURL      \undefined \def \showURL       {\relax}        \fi
\providecommand\bibfield[2]{#2}
\providecommand\bibinfo[2]{#2}
\providecommand\natexlab[1]{#1}
\providecommand\showeprint[2][]{arXiv:#2}

\bibitem[Alayrac et~al\mbox{.}(2022)]%
        {flamingo}
\bibfield{author}{\bibinfo{person}{Jean-Baptiste Alayrac}, \bibinfo{person}{Jeff Donahue}, \bibinfo{person}{Pauline Luc}, \bibinfo{person}{Antoine Miech}, \bibinfo{person}{Iain Barr}, \bibinfo{person}{Yana Hasson}, \bibinfo{person}{Karel Lenc}, \bibinfo{person}{Arthur Mensch}, \bibinfo{person}{Katherine Millican}, \bibinfo{person}{Malcolm Reynolds}, {et~al\mbox{.}}} \bibinfo{year}{2022}\natexlab{}.
\newblock \showarticletitle{Flamingo: a visual language model for few-shot learning}.
\newblock \bibinfo{journal}{\emph{Advances in Neural Information Processing Systems (NeurIPS)}} (\bibinfo{year}{2022}), \bibinfo{pages}{23716--23736}.
\newblock


\bibitem[Ao et~al\mbox{.}(2023)]%
        {PNG}
\bibfield{author}{\bibinfo{person}{Xiang Ao}, \bibinfo{person}{Ling Luo}, \bibinfo{person}{Xiting Wang}, \bibinfo{person}{Zhao Yang}, \bibinfo{person}{Jiun-Hung Chen}, \bibinfo{person}{Ying Qiao}, \bibinfo{person}{Qing He}, {and} \bibinfo{person}{Xing Xie}.} \bibinfo{year}{2023}\natexlab{}.
\newblock \showarticletitle{Put Your Voice on Stage: Personalized Headline Generation for News Articles}.
\newblock \bibinfo{journal}{\emph{TKDD}} \bibinfo{volume}{18}, \bibinfo{number}{3} (\bibinfo{year}{2023}).
\newblock


\bibitem[Ao et~al\mbox{.}(2021)]%
        {PENS}
\bibfield{author}{\bibinfo{person}{Xiang Ao}, \bibinfo{person}{Xiting Wang}, \bibinfo{person}{Ling Luo}, \bibinfo{person}{Ying Qiao}, \bibinfo{person}{Qing He}, {and} \bibinfo{person}{Xing Xie}.} \bibinfo{year}{2021}\natexlab{}.
\newblock \showarticletitle{{PENS:} {A} Dataset and Generic Framework for Personalized News Headline Generation}. In \bibinfo{booktitle}{\emph{Proceedings of ACL/IJCNLP}}. \bibinfo{pages}{82--92}.
\newblock


\bibitem[Baltescu et~al\mbox{.}(2022)]%
        {itemsage}
\bibfield{author}{\bibinfo{person}{Paul Baltescu}, \bibinfo{person}{Haoyu Chen}, \bibinfo{person}{Nikil Pancha}, \bibinfo{person}{Andrew Zhai}, \bibinfo{person}{Jure Leskovec}, {and} \bibinfo{person}{Charles Rosenberg}.} \bibinfo{year}{2022}\natexlab{}.
\newblock \showarticletitle{Itemsage: Learning product embeddings for shopping recommendations at pinterest}. In \bibinfo{booktitle}{\emph{Proceedings of the 28th ACM SIGKDD Conference on Knowledge Discovery and Data Mining (KDD)}}. \bibinfo{pages}{2703--2711}.
\newblock


\bibitem[Bank et~al\mbox{.}(2020)]%
        {ae}
\bibfield{author}{\bibinfo{person}{Dor Bank}, \bibinfo{person}{Noam Koenigstein}, {and} \bibinfo{person}{Raja Giryes}.} \bibinfo{year}{2020}\natexlab{}.
\newblock \showarticletitle{Autoencoders}.
\newblock \bibinfo{journal}{\emph{CoRR}}  \bibinfo{volume}{abs/2003.05991} (\bibinfo{year}{2020}).
\newblock


\bibitem[Bian et~al\mbox{.}(2023)]%
        {M3SRec}
\bibfield{author}{\bibinfo{person}{Shuqing Bian}, \bibinfo{person}{Xingyu Pan}, \bibinfo{person}{Wayne~Xin Zhao}, \bibinfo{person}{Jinpeng Wang}, \bibinfo{person}{Chuyuan Wang}, {and} \bibinfo{person}{Ji{-}Rong Wen}.} \bibinfo{year}{2023}\natexlab{}.
\newblock \showarticletitle{Multi-modal Mixture of Experts Represetation Learning for Sequential Recommendation}. In \bibinfo{booktitle}{\emph{Proceedings of the 32nd {ACM} International Conference on Information and Knowledge Management (CIKM)}}. \bibinfo{pages}{110--119}.
\newblock


\bibitem[Brown et~al\mbox{.}(2020)]%
        {gpt-3}
\bibfield{author}{\bibinfo{person}{Tom Brown}, \bibinfo{person}{Benjamin Mann}, \bibinfo{person}{Nick Ryder}, \bibinfo{person}{Melanie Subbiah}, \bibinfo{person}{Jared~D Kaplan}, \bibinfo{person}{Prafulla Dhariwal}, \bibinfo{person}{Arvind Neelakantan}, \bibinfo{person}{Pranav Shyam}, \bibinfo{person}{Girish Sastry}, \bibinfo{person}{Amanda Askell}, {et~al\mbox{.}}} \bibinfo{year}{2020}\natexlab{}.
\newblock \showarticletitle{Language models are few-shot learners}.
\newblock \bibinfo{journal}{\emph{Advances in Neural Information Processing Systems (NeurIPS)}} (\bibinfo{year}{2020}), \bibinfo{pages}{1877--1901}.
\newblock


\bibitem[Cai et~al\mbox{.}(2023)]%
        {GUE}
\bibfield{author}{\bibinfo{person}{Pengshan Cai}, \bibinfo{person}{Kaiqiang Song}, \bibinfo{person}{Sangwoo Cho}, \bibinfo{person}{Hongwei Wang}, \bibinfo{person}{Xiaoyang Wang}, \bibinfo{person}{Hong Yu}, \bibinfo{person}{Fei Liu}, {and} \bibinfo{person}{Dong Yu}.} \bibinfo{year}{2023}\natexlab{}.
\newblock \showarticletitle{Generating User-Engaging News Headlines}. In \bibinfo{booktitle}{\emph{Proceedings of ACL}}. \bibinfo{pages}{3265--3280}.
\newblock


\bibitem[Chen et~al\mbox{.}(2021b)]%
        {AutomatedCreative}
\bibfield{author}{\bibinfo{person}{Jin Chen}, \bibinfo{person}{Ju Xu}, \bibinfo{person}{Gangwei Jiang}, \bibinfo{person}{Tiezheng Ge}, \bibinfo{person}{Zhiqiang Zhang}, \bibinfo{person}{Defu Lian}, {and} \bibinfo{person}{Kai Zheng}.} \bibinfo{year}{2021}\natexlab{b}.
\newblock \showarticletitle{Automated Creative Optimization for E-Commerce Advertising}. In \bibinfo{booktitle}{\emph{The ACM Web Conference ({WWW})}}. \bibinfo{pages}{2304--2313}.
\newblock


\bibitem[Chen et~al\mbox{.}(2021a)]%
        {uae}
\bibfield{author}{\bibinfo{person}{Ke Chen}, \bibinfo{person}{Beici Liang}, \bibinfo{person}{Xiaoshuan Ma}, {and} \bibinfo{person}{Minwei Gu}.} \bibinfo{year}{2021}\natexlab{a}.
\newblock \showarticletitle{Learning audio embeddings with user listening data for content-based music recommendation}. In \bibinfo{booktitle}{\emph{IEEE International Conference on Acoustics, Speech and Signal Processing (ICASSP)}}. IEEE, \bibinfo{pages}{3015--3019}.
\newblock


\bibitem[Chen et~al\mbox{.}(2020)]%
        {SimCLR}
\bibfield{author}{\bibinfo{person}{Ting Chen}, \bibinfo{person}{Simon Kornblith}, \bibinfo{person}{Mohammad Norouzi}, {and} \bibinfo{person}{Geoffrey~E. Hinton}.} \bibinfo{year}{2020}\natexlab{}.
\newblock \showarticletitle{A Simple Framework for Contrastive Learning of Visual Representations}. In \bibinfo{booktitle}{\emph{Proceedings of the 37th International Conference on Machine Learning (ICML)}}. \bibinfo{pages}{1597--1607}.
\newblock


\bibitem[Chen et~al\mbox{.}(2022)]%
        {hccm}
\bibfield{author}{\bibinfo{person}{Xin Chen}, \bibinfo{person}{Qingtao Tang}, \bibinfo{person}{Ke Hu}, \bibinfo{person}{Yue Xu}, \bibinfo{person}{Shihang Qiu}, \bibinfo{person}{Jia Cheng}, {and} \bibinfo{person}{Jun Lei}.} \bibinfo{year}{2022}\natexlab{}.
\newblock \showarticletitle{Hybrid CNN Based Attention with Category Prior for User Image Behavior Modeling}. In \bibinfo{booktitle}{\emph{Proceedings of the 45th International ACM SIGIR Conference on Research and Development in Information Retrieval (SIGIR)}}. \bibinfo{pages}{2336--2340}.
\newblock


\bibitem[Deldjoo et~al\mbox{.}(2024)]%
        {DeldjooNRMPBN24}
\bibfield{author}{\bibinfo{person}{Yashar Deldjoo}, \bibinfo{person}{Fatemeh Nazary}, \bibinfo{person}{Arnau Ramisa}, \bibinfo{person}{Julian~J. McAuley}, \bibinfo{person}{Giovanni Pellegrini}, \bibinfo{person}{Alejandro Bellog{\'{\i}}n}, {and} \bibinfo{person}{Tommaso~Di Noia}.} \bibinfo{year}{2024}\natexlab{}.
\newblock \showarticletitle{A Review of Modern Fashion Recommender Systems}.
\newblock \bibinfo{journal}{\emph{{ACM} Comput. Surv.}} \bibinfo{volume}{56}, \bibinfo{number}{4} (\bibinfo{year}{2024}), \bibinfo{pages}{87:1--87:37}.
\newblock


\bibitem[Deldjoo et~al\mbox{.}(2020)]%
        {DeldjooSCP20}
\bibfield{author}{\bibinfo{person}{Yashar Deldjoo}, \bibinfo{person}{Markus Schedl}, \bibinfo{person}{Paolo Cremonesi}, {and} \bibinfo{person}{Gabriella Pasi}.} \bibinfo{year}{2020}\natexlab{}.
\newblock \showarticletitle{Recommender systems leveraging multimedia content}.
\newblock \bibinfo{journal}{\emph{Comput. Surveys}} \bibinfo{volume}{53}, \bibinfo{number}{5} (\bibinfo{year}{2020}), \bibinfo{pages}{1--38}.
\newblock


\bibitem[Deldjoo et~al\mbox{.}(2021)]%
        {MusicSurvey}
\bibfield{author}{\bibinfo{person}{Yashar Deldjoo}, \bibinfo{person}{Markus Schedl}, {and} \bibinfo{person}{Peter Knees}.} \bibinfo{year}{2021}\natexlab{}.
\newblock \showarticletitle{Content-driven Music Recommendation: Evolution, State of the Art, and Challenges}.
\newblock \bibinfo{journal}{\emph{CoRR}}  \bibinfo{volume}{abs/2107.11803} (\bibinfo{year}{2021}).
\newblock


\bibitem[Deng et~al\mbox{.}(2024)]%
        {EM3}
\bibfield{author}{\bibinfo{person}{Xiuqi Deng}, \bibinfo{person}{Lu Xu}, \bibinfo{person}{Xiyao Li}, \bibinfo{person}{Jinkai Yu}, \bibinfo{person}{Erpeng Xue}, \bibinfo{person}{Zhongyuan Wang}, \bibinfo{person}{Di Zhang}, \bibinfo{person}{Zhaojie Liu}, \bibinfo{person}{Guorui Zhou}, \bibinfo{person}{Yang Song}, \bibinfo{person}{Na Mou}, \bibinfo{person}{Shen Jiang}, {and} \bibinfo{person}{Han Li}.} \bibinfo{year}{2024}\natexlab{}.
\newblock \showarticletitle{End-to-end training of Multimodal Model and ranking Model}.
\newblock \bibinfo{journal}{\emph{CoRR}}  \bibinfo{volume}{abs/2404.06078} (\bibinfo{year}{2024}).
\newblock


\bibitem[Deng et~al\mbox{.}(2022)]%
        {DengLZDL22}
\bibfield{author}{\bibinfo{person}{Yang Deng}, \bibinfo{person}{Yaliang Li}, \bibinfo{person}{Wenxuan Zhang}, \bibinfo{person}{Bolin Ding}, {and} \bibinfo{person}{Wai Lam}.} \bibinfo{year}{2022}\natexlab{}.
\newblock \showarticletitle{Toward Personalized Answer Generation in E-Commerce via Multi-perspective Preference Modeling}.
\newblock \bibinfo{journal}{\emph{{ACM} Trans. Inf. Syst.}} \bibinfo{volume}{40}, \bibinfo{number}{4} (\bibinfo{year}{2022}), \bibinfo{pages}{87:1--87:28}.
\newblock


\bibitem[Devlin et~al\mbox{.}(2019)]%
        {bert}
\bibfield{author}{\bibinfo{person}{Jacob Devlin}, \bibinfo{person}{Ming{-}Wei Chang}, \bibinfo{person}{Kenton Lee}, {and} \bibinfo{person}{Kristina Toutanova}.} \bibinfo{year}{2019}\natexlab{}.
\newblock \showarticletitle{{BERT:} Pre-training of Deep Bidirectional Transformers for Language Understanding}. In \bibinfo{booktitle}{\emph{Proceedings of the 2019 Conference of the North American Chapter of the Association for Computational Linguistics: Human Language Technologies ({NAACL-HLT})}}. \bibinfo{pages}{4171--4186}.
\newblock


\bibitem[Ding et~al\mbox{.}(2023)]%
        {DingSZTJ23}
\bibfield{author}{\bibinfo{person}{Zijian Ding}, \bibinfo{person}{Alison Smith{-}Renner}, \bibinfo{person}{Wenjuan Zhang}, \bibinfo{person}{Joel~R. Tetreault}, {and} \bibinfo{person}{Alejandro Jaimes}.} \bibinfo{year}{2023}\natexlab{}.
\newblock \showarticletitle{Harnessing the power of LLMs: Evaluating human-AI text co-creation through the lens of news headline generation}. In \bibinfo{booktitle}{\emph{Findings of {EMNLP}}}. \bibinfo{pages}{3321--3339}.
\newblock


\bibitem[Dong et~al\mbox{.}(2022)]%
        {M5Product}
\bibfield{author}{\bibinfo{person}{Xiao Dong}, \bibinfo{person}{Xunlin Zhan}, \bibinfo{person}{Yangxin Wu}, \bibinfo{person}{Yunchao Wei}, \bibinfo{person}{Michael~C. Kampffmeyer}, \bibinfo{person}{Xiaoyong Wei}, \bibinfo{person}{Minlong Lu}, \bibinfo{person}{Yaowei Wang}, {and} \bibinfo{person}{Xiaodan Liang}.} \bibinfo{year}{2022}\natexlab{}.
\newblock \showarticletitle{M5Product: Self-harmonized Contrastive Learning for E-commercial Multi-modal Pretraining}. In \bibinfo{booktitle}{\emph{{IEEE/CVF} Conference on Computer Vision and Pattern Recognition (CVPR)}}. \bibinfo{pages}{21220--21230}.
\newblock


\bibitem[Dosovitskiy et~al\mbox{.}(2021)]%
        {vit}
\bibfield{author}{\bibinfo{person}{Alexey Dosovitskiy}, \bibinfo{person}{Lucas Beyer}, \bibinfo{person}{Alexander Kolesnikov}, \bibinfo{person}{Dirk Weissenborn}, \bibinfo{person}{Xiaohua Zhai}, \bibinfo{person}{Thomas Unterthiner}, \bibinfo{person}{Mostafa Dehghani}, \bibinfo{person}{Matthias Minderer}, \bibinfo{person}{Georg Heigold}, \bibinfo{person}{Sylvain Gelly}, \bibinfo{person}{Jakob Uszkoreit}, {and} \bibinfo{person}{Neil Houlsby}.} \bibinfo{year}{2021}\natexlab{}.
\newblock \showarticletitle{An Image is Worth 16x16 Words: Transformers for Image Recognition at Scale}. In \bibinfo{booktitle}{\emph{9th International Conference on Learning Representations (ICLR)}}.
\newblock


\bibitem[Du et~al\mbox{.}(2020)]%
        {mtpr}
\bibfield{author}{\bibinfo{person}{Xiaoyu Du}, \bibinfo{person}{Xiang Wang}, \bibinfo{person}{Xiangnan He}, \bibinfo{person}{Zechao Li}, \bibinfo{person}{Jinhui Tang}, {and} \bibinfo{person}{Tat-Seng Chua}.} \bibinfo{year}{2020}\natexlab{}.
\newblock \showarticletitle{How to learn item representation for cold-start multimedia recommendation?}. In \bibinfo{booktitle}{\emph{Proceedings of the 28th ACM International Conference on Multimedia}}. \bibinfo{pages}{3469--3477}.
\newblock


\bibitem[Duan et~al\mbox{.}(2023)]%
        {Crossmodalprompt}
\bibfield{author}{\bibinfo{person}{Haoyi Duan}, \bibinfo{person}{Yan Xia}, \bibinfo{person}{Mingze Zhou}, \bibinfo{person}{Li Tang}, \bibinfo{person}{Jieming Zhu}, {and} \bibinfo{person}{Zhou Zhao}.} \bibinfo{year}{2023}\natexlab{}.
\newblock \showarticletitle{Cross-modal Prompts: Adapting Large Pre-trained Models for Audio-Visual Downstream Tasks}. In \bibinfo{booktitle}{\emph{Advances in Neural Information Processing Systems (NeurIPS)}}.
\newblock


\bibitem[Elizalde et~al\mbox{.}(2023)]%
        {clap}
\bibfield{author}{\bibinfo{person}{Benjamin Elizalde}, \bibinfo{person}{Soham Deshmukh}, \bibinfo{person}{Mahmoud Al~Ismail}, {and} \bibinfo{person}{Huaming Wang}.} \bibinfo{year}{2023}\natexlab{}.
\newblock \showarticletitle{Clap learning audio concepts from natural language supervision}. In \bibinfo{booktitle}{\emph{IEEE International Conference on Acoustics, Speech and Signal Processing (ICASSP)}}. IEEE, \bibinfo{pages}{1--5}.
\newblock


\bibitem[Fang et~al\mbox{.}(2024)]%
        {fang2024multi}
\bibfield{author}{\bibinfo{person}{Jiabao Fang}, \bibinfo{person}{Shen Gao}, \bibinfo{person}{Pengjie Ren}, \bibinfo{person}{Xiuying Chen}, \bibinfo{person}{Suzan Verberne}, {and} \bibinfo{person}{Zhaochun Ren}.} \bibinfo{year}{2024}\natexlab{}.
\newblock \showarticletitle{A Multi-Agent Conversational Recommender System}.
\newblock \bibinfo{journal}{\emph{CoRR}}  \bibinfo{volume}{abs/2402.01135} (\bibinfo{year}{2024}).
\newblock


\bibitem[Feng et~al\mbox{.}(2023)]%
        {LLMCRS}
\bibfield{author}{\bibinfo{person}{Yue Feng}, \bibinfo{person}{Shuchang Liu}, \bibinfo{person}{Zhenghai Xue}, \bibinfo{person}{Qingpeng Cai}, \bibinfo{person}{Lantao Hu}, \bibinfo{person}{Peng Jiang}, \bibinfo{person}{Kun Gai}, {and} \bibinfo{person}{Fei Sun}.} \bibinfo{year}{2023}\natexlab{}.
\newblock \showarticletitle{A Large Language Model Enhanced Conversational Recommender System}.
\newblock \bibinfo{journal}{\emph{CoRR}}  \bibinfo{volume}{abs/2308.06212} (\bibinfo{year}{2023}).
\newblock


\bibitem[Fu et~al\mbox{.}(2024)]%
        {transrec}
\bibfield{author}{\bibinfo{person}{Junchen Fu}, \bibinfo{person}{Fajie Yuan}, \bibinfo{person}{Yu Song}, \bibinfo{person}{Zheng Yuan}, \bibinfo{person}{Mingyue Cheng}, \bibinfo{person}{Shenghui Cheng}, \bibinfo{person}{Jiaqi Zhang}, \bibinfo{person}{Jie Wang}, {and} \bibinfo{person}{Yunzhu Pan}.} \bibinfo{year}{2024}\natexlab{}.
\newblock \showarticletitle{Exploring adapter-based transfer learning for recommender systems: Empirical studies and practical insights}. In \bibinfo{booktitle}{\emph{Proceedings of the 17th ACM International Conference on Web Search and Data Mining (WSDM)}}. \bibinfo{pages}{208--217}.
\newblock


\bibitem[Gao et~al\mbox{.}(2021)]%
        {SimCSE}
\bibfield{author}{\bibinfo{person}{Tianyu Gao}, \bibinfo{person}{Xingcheng Yao}, {and} \bibinfo{person}{Danqi Chen}.} \bibinfo{year}{2021}\natexlab{}.
\newblock \showarticletitle{SimCSE: Simple Contrastive Learning of Sentence Embeddings}. In \bibinfo{booktitle}{\emph{Proceedings of the 2021 Conference on Empirical Methods in Natural Language Processing (EMNLP)}}. \bibinfo{pages}{6894--6910}.
\newblock


\bibitem[Gao et~al\mbox{.}(2023)]%
        {TextPainter}
\bibfield{author}{\bibinfo{person}{Yifan Gao}, \bibinfo{person}{Jinpeng Lin}, \bibinfo{person}{Min Zhou}, \bibinfo{person}{Chuanbin Liu}, \bibinfo{person}{Hongtao Xie}, \bibinfo{person}{Tiezheng Ge}, {and} \bibinfo{person}{Yuning Jiang}.} \bibinfo{year}{2023}\natexlab{}.
\newblock \showarticletitle{TextPainter: Multimodal Text Image Generation with Visual-harmony and Text-comprehension for Poster Design}. In \bibinfo{booktitle}{\emph{ACM MM}}. \bibinfo{pages}{7236--7246}.
\newblock


\bibitem[Ge et~al\mbox{.}(2018)]%
        {ImageMatters}
\bibfield{author}{\bibinfo{person}{Tiezheng Ge}, \bibinfo{person}{Liqin Zhao}, \bibinfo{person}{Guorui Zhou}, \bibinfo{person}{Keyu Chen}, \bibinfo{person}{Shuying Liu}, \bibinfo{person}{Huiming Yi}, \bibinfo{person}{Zelin Hu}, \bibinfo{person}{Bochao Liu}, \bibinfo{person}{Peng Sun}, \bibinfo{person}{Haoyu Liu}, \bibinfo{person}{Pengtao Yi}, \bibinfo{person}{Sui Huang}, \bibinfo{person}{Zhiqiang Zhang}, \bibinfo{person}{Xiaoqiang Zhu}, \bibinfo{person}{Yu Zhang}, {and} \bibinfo{person}{Kun Gai}.} \bibinfo{year}{2018}\natexlab{}.
\newblock \showarticletitle{Image Matters: Visually Modeling User Behaviors Using Advanced Model Server}. In \bibinfo{booktitle}{\emph{CIKM}}. \bibinfo{pages}{2087--2095}.
\newblock


\bibitem[Geng et~al\mbox{.}(2022)]%
        {p5}
\bibfield{author}{\bibinfo{person}{Shijie Geng}, \bibinfo{person}{Shuchang Liu}, \bibinfo{person}{Zuohui Fu}, \bibinfo{person}{Yingqiang Ge}, {and} \bibinfo{person}{Yongfeng Zhang}.} \bibinfo{year}{2022}\natexlab{}.
\newblock \showarticletitle{Recommendation as language processing (rlp): A unified pretrain, personalized prompt \& predict paradigm (p5)}. In \bibinfo{booktitle}{\emph{Proceedings of the 16th ACM Conference on Recommender Systems (RecSys)}}. \bibinfo{pages}{299--315}.
\newblock


\bibitem[Geng et~al\mbox{.}(2023)]%
        {vip5}
\bibfield{author}{\bibinfo{person}{Shijie Geng}, \bibinfo{person}{Juntao Tan}, \bibinfo{person}{Shuchang Liu}, \bibinfo{person}{Zuohui Fu}, {and} \bibinfo{person}{Yongfeng Zhang}.} \bibinfo{year}{2023}\natexlab{}.
\newblock \showarticletitle{VIP5: Towards Multimodal Foundation Models for Recommendation}. In \bibinfo{booktitle}{\emph{Findings of the Association for Computational Linguistics: EMNLP 2023}}. \bibinfo{pages}{9606--9620}.
\newblock


\bibitem[Girdhar et~al\mbox{.}(2023)]%
        {imagebind}
\bibfield{author}{\bibinfo{person}{Rohit Girdhar}, \bibinfo{person}{Alaaeldin El-Nouby}, \bibinfo{person}{Zhuang Liu}, \bibinfo{person}{Mannat Singh}, \bibinfo{person}{Kalyan~Vasudev Alwala}, \bibinfo{person}{Armand Joulin}, {and} \bibinfo{person}{Ishan Misra}.} \bibinfo{year}{2023}\natexlab{}.
\newblock \showarticletitle{Imagebind: One embedding space to bind them all}. In \bibinfo{booktitle}{\emph{Proceedings of the IEEE/CVF Conference on Computer Vision and Pattern Recognition (CVPR)}}. \bibinfo{pages}{15180--15190}.
\newblock


\bibitem[Gong et~al\mbox{.}(2024)]%
        {AtomoVideo}
\bibfield{author}{\bibinfo{person}{Litong Gong}, \bibinfo{person}{Yiran Zhu}, \bibinfo{person}{Weijie Li}, \bibinfo{person}{Xiaoyang Kang}, \bibinfo{person}{Biao Wang}, \bibinfo{person}{Tiezheng Ge}, {and} \bibinfo{person}{Bo Zheng}.} \bibinfo{year}{2024}\natexlab{}.
\newblock \showarticletitle{AtomoVideo: High Fidelity Image-to-Video Generation}.
\newblock  (\bibinfo{year}{2024}).
\newblock
\showeprint{2403.01800}


\bibitem[Gu et~al\mbox{.}(2020)]%
        {NHNet}
\bibfield{author}{\bibinfo{person}{Xiaotao Gu}, \bibinfo{person}{Yuning Mao}, \bibinfo{person}{Jiawei Han}, \bibinfo{person}{Jialu Liu}, \bibinfo{person}{You Wu}, \bibinfo{person}{Cong Yu}, \bibinfo{person}{Daniel Finnie}, \bibinfo{person}{Hongkun Yu}, \bibinfo{person}{Jiaqi Zhai}, {and} \bibinfo{person}{Nicholas Zukoski}.} \bibinfo{year}{2020}\natexlab{}.
\newblock \showarticletitle{Generating Representative Headlines for News Stories}. In \bibinfo{booktitle}{\emph{The Web Conference 2020 (WWW)}}. \bibinfo{pages}{1773--1784}.
\newblock


\bibitem[He et~al\mbox{.}(2022)]%
        {mae}
\bibfield{author}{\bibinfo{person}{Kaiming He}, \bibinfo{person}{Xinlei Chen}, \bibinfo{person}{Saining Xie}, \bibinfo{person}{Yanghao Li}, \bibinfo{person}{Piotr Doll{\'a}r}, {and} \bibinfo{person}{Ross Girshick}.} \bibinfo{year}{2022}\natexlab{}.
\newblock \showarticletitle{Masked autoencoders are scalable vision learners}. In \bibinfo{booktitle}{\emph{Proceedings of the IEEE/CVF Conference on Computer Vision and Pattern Recognition (CVPR)}}. \bibinfo{pages}{16000--16009}.
\newblock


\bibitem[He et~al\mbox{.}(2016)]%
        {resnet}
\bibfield{author}{\bibinfo{person}{Kaiming He}, \bibinfo{person}{Xiangyu Zhang}, \bibinfo{person}{Shaoqing Ren}, {and} \bibinfo{person}{Jian Sun}.} \bibinfo{year}{2016}\natexlab{}.
\newblock \showarticletitle{Deep residual learning for image recognition}. In \bibinfo{booktitle}{\emph{Proceedings of the IEEE/CVF Conference on Computer Vision and Pattern Recognition (CVPR)}}. \bibinfo{pages}{770--778}.
\newblock


\bibitem[Hou et~al\mbox{.}(2022)]%
        {unisrec}
\bibfield{author}{\bibinfo{person}{Yupeng Hou}, \bibinfo{person}{Shanlei Mu}, \bibinfo{person}{Wayne~Xin Zhao}, \bibinfo{person}{Yaliang Li}, \bibinfo{person}{Bolin Ding}, {and} \bibinfo{person}{Ji-Rong Wen}.} \bibinfo{year}{2022}\natexlab{}.
\newblock \showarticletitle{Towards universal sequence representation learning for recommender systems}. In \bibinfo{booktitle}{\emph{Proceedings of the 28th ACM SIGKDD Conference on Knowledge Discovery and Data Mining (KDD)}}. \bibinfo{pages}{585--593}.
\newblock


\bibitem[Hsu et~al\mbox{.}(2023)]%
        {PosterLayout}
\bibfield{author}{\bibinfo{person}{HsiaoYuan Hsu}, \bibinfo{person}{Xiangteng He}, \bibinfo{person}{Yuxin Peng}, \bibinfo{person}{Hao Kong}, {and} \bibinfo{person}{Qing Zhang}.} \bibinfo{year}{2023}\natexlab{}.
\newblock \showarticletitle{PosterLayout: {A} New Benchmark and Approach for Content-Aware Visual-Textual Presentation Layout}. In \bibinfo{booktitle}{\emph{CVPR}}. \bibinfo{pages}{6018--6026}.
\newblock


\bibitem[Hu et~al\mbox{.}(2022)]%
        {lora}
\bibfield{author}{\bibinfo{person}{Edward~J. Hu}, \bibinfo{person}{Yelong Shen}, \bibinfo{person}{Phillip Wallis}, \bibinfo{person}{Zeyuan Allen{-}Zhu}, \bibinfo{person}{Yuanzhi Li}, \bibinfo{person}{Shean Wang}, \bibinfo{person}{Lu Wang}, {and} \bibinfo{person}{Weizhu Chen}.} \bibinfo{year}{2022}\natexlab{}.
\newblock \showarticletitle{LoRA: Low-Rank Adaptation of Large Language Models}. In \bibinfo{booktitle}{\emph{The Tenth International Conference on Learning Representations, {ICLR} 2022, Virtual Event, April 25-29, 2022}}.
\newblock


\bibitem[Hu et~al\mbox{.}(2024)]%
        {kdsr}
\bibfield{author}{\bibinfo{person}{Hengchang Hu}, \bibinfo{person}{Qijiong Liu}, \bibinfo{person}{Chuang Li}, {and} \bibinfo{person}{Min-Yen Kan}.} \bibinfo{year}{2024}\natexlab{}.
\newblock \showarticletitle{Lightweight Modality Adaptation to Sequential Recommendation via Correlation Supervision}.
\newblock \bibinfo{journal}{\emph{arXiv preprint arXiv:2401.07257}} (\bibinfo{year}{2024}).
\newblock


\bibitem[Huang et~al\mbox{.}(2024)]%
        {FoundationSurvey}
\bibfield{author}{\bibinfo{person}{Chengkai Huang}, \bibinfo{person}{Tong Yu}, \bibinfo{person}{Kaige Xie}, \bibinfo{person}{Shuai Zhang}, \bibinfo{person}{Lina Yao}, {and} \bibinfo{person}{Julian~J. McAuley}.} \bibinfo{year}{2024}\natexlab{}.
\newblock \showarticletitle{Foundation Models for Recommender Systems: {A} Survey and New Perspectives}.
\newblock \bibinfo{journal}{\emph{CoRR}}  \bibinfo{volume}{abs/2402.11143} (\bibinfo{year}{2024}).
\newblock


\bibitem[Huang et~al\mbox{.}(2020)]%
        {huang2020large}
\bibfield{author}{\bibinfo{person}{Qingqing Huang}, \bibinfo{person}{Aren Jansen}, \bibinfo{person}{Li Zhang}, \bibinfo{person}{Daniel~PW Ellis}, \bibinfo{person}{Rif~A Saurous}, {and} \bibinfo{person}{John Anderson}.} \bibinfo{year}{2020}\natexlab{}.
\newblock \showarticletitle{Large-scale weakly-supervised content embeddings for music recommendation and tagging}. In \bibinfo{booktitle}{\emph{IEEE International Conference on Acoustics, Speech and Signal Processing (ICASSP)}}. \bibinfo{pages}{8364--8368}.
\newblock


\bibitem[Huang et~al\mbox{.}(2021)]%
        {ssd}
\bibfield{author}{\bibinfo{person}{Yanhua Huang}, \bibinfo{person}{Weikun Wang}, \bibinfo{person}{Lei Zhang}, {and} \bibinfo{person}{Ruiwen Xu}.} \bibinfo{year}{2021}\natexlab{}.
\newblock \showarticletitle{Sliding spectrum decomposition for diversified recommendation}. In \bibinfo{booktitle}{\emph{Proceedings of the 27th ACM SIGKDD Conference on Knowledge Discovery \& Data Mining (KDD)}}. \bibinfo{pages}{3041--3049}.
\newblock


\bibitem[Inoue et~al\mbox{.}(2023)]%
        {LayoutDM}
\bibfield{author}{\bibinfo{person}{Naoto Inoue}, \bibinfo{person}{Kotaro Kikuchi}, \bibinfo{person}{Edgar Simo{-}Serra}, \bibinfo{person}{Mayu Otani}, {and} \bibinfo{person}{Kota Yamaguchi}.} \bibinfo{year}{2023}\natexlab{}.
\newblock \showarticletitle{LayoutDM: Discrete Diffusion Model for Controllable Layout Generation}. In \bibinfo{booktitle}{\emph{CVPR}}. \bibinfo{pages}{10167--10176}.
\newblock


\bibitem[Jin et~al\mbox{.}(2024)]%
        {CoST}
\bibfield{author}{\bibinfo{person}{Mengqun Jin}, \bibinfo{person}{Zexuan Qiu}, \bibinfo{person}{Jieming Zhu}, \bibinfo{person}{Zhenhua Dong}, {and} \bibinfo{person}{Xiu Li}.} \bibinfo{year}{2024}\natexlab{}.
\newblock \showarticletitle{Contrastive Quantization based Semantic Code for Generative Recommendation}.
\newblock \bibinfo{journal}{\emph{CoRR}}  \bibinfo{volume}{abs/2404.14774} (\bibinfo{year}{2024}).
\newblock


\bibitem[Jin et~al\mbox{.}(2023)]%
        {ECLIP}
\bibfield{author}{\bibinfo{person}{Yang Jin}, \bibinfo{person}{Yongzhi Li}, \bibinfo{person}{Zehuan Yuan}, {and} \bibinfo{person}{Yadong Mu}.} \bibinfo{year}{2023}\natexlab{}.
\newblock \showarticletitle{Learning Instance-Level Representation for Large-Scale Multi-Modal Pretraining in E-Commerce}. In \bibinfo{booktitle}{\emph{{IEEE/CVF} Conference on Computer Vision and Pattern Recognition (CVPR)}}. \bibinfo{pages}{11060--11069}.
\newblock


\bibitem[Kang and McAuley(2018)]%
        {sasrec}
\bibfield{author}{\bibinfo{person}{Wang-Cheng Kang} {and} \bibinfo{person}{Julian McAuley}.} \bibinfo{year}{2018}\natexlab{}.
\newblock \showarticletitle{Self-attentive sequential recommendation}. In \bibinfo{booktitle}{\emph{IEEE International Conference on Data Mining (ICDM)}}. IEEE, \bibinfo{pages}{197--206}.
\newblock


\bibitem[Khattak et~al\mbox{.}(2023)]%
        {khattak2023maple}
\bibfield{author}{\bibinfo{person}{Muhammad~Uzair Khattak}, \bibinfo{person}{Hanoona~Abdul Rasheed}, \bibinfo{person}{Muhammad Maaz}, \bibinfo{person}{Salman~H. Khan}, {and} \bibinfo{person}{Fahad~Shahbaz Khan}.} \bibinfo{year}{2023}\natexlab{}.
\newblock \showarticletitle{MaPLe: Multi-modal Prompt Learning}. In \bibinfo{booktitle}{\emph{{IEEE/CVF} Conference on Computer Vision and Pattern Recognition (CVPR)}}. \bibinfo{pages}{19113--19122}.
\newblock


\bibitem[Kingma and Welling(2014)]%
        {vae}
\bibfield{author}{\bibinfo{person}{Diederik~P. Kingma} {and} \bibinfo{person}{Max Welling}.} \bibinfo{year}{2014}\natexlab{}.
\newblock \showarticletitle{Auto-Encoding Variational Bayes}. In \bibinfo{booktitle}{\emph{2nd International Conference on Learning Representations (ICLR)}}, \bibfield{editor}{\bibinfo{person}{Yoshua Bengio} {and} \bibinfo{person}{Yann LeCun}} (Eds.).
\newblock


\bibitem[Krubinski and Pecina(2024)]%
        {NewsGen}
\bibfield{author}{\bibinfo{person}{Mateusz Krubinski} {and} \bibinfo{person}{Pavel Pecina}.} \bibinfo{year}{2024}\natexlab{}.
\newblock \showarticletitle{Towards Unified Uni- and Multi-modal News Headline Generation}. In \bibinfo{booktitle}{\emph{Proceedings of {EACL}}}. \bibinfo{pages}{437--450}.
\newblock


\bibitem[Lewis et~al\mbox{.}(2020)]%
        {bart}
\bibfield{author}{\bibinfo{person}{Mike Lewis}, \bibinfo{person}{Yinhan Liu}, \bibinfo{person}{Naman Goyal}, \bibinfo{person}{Marjan Ghazvininejad}, \bibinfo{person}{Abdelrahman Mohamed}, \bibinfo{person}{Omer Levy}, \bibinfo{person}{Veselin Stoyanov}, {and} \bibinfo{person}{Luke Zettlemoyer}.} \bibinfo{year}{2020}\natexlab{}.
\newblock \showarticletitle{{BART:} Denoising Sequence-to-Sequence Pre-training for Natural Language Generation, Translation, and Comprehension}. In \bibinfo{booktitle}{\emph{Proceedings of the 58th Annual Meeting of the Association for Computational Linguistics (ACL)}}. \bibinfo{pages}{7871--7880}.
\newblock


\bibitem[Li et~al\mbox{.}(2023a)]%
        {TagGPT}
\bibfield{author}{\bibinfo{person}{Chen Li}, \bibinfo{person}{Yixiao Ge}, \bibinfo{person}{Jiayong Mao}, \bibinfo{person}{Dian Li}, {and} \bibinfo{person}{Ying Shan}.} \bibinfo{year}{2023}\natexlab{a}.
\newblock \showarticletitle{TagGPT: Large Language Models are Zero-shot Multimodal Taggers}.
\newblock \bibinfo{journal}{\emph{CoRR}}  \bibinfo{volume}{abs/2304.03022} (\bibinfo{year}{2023}).
\newblock


\bibitem[Li et~al\mbox{.}(2023b)]%
        {blip-2}
\bibfield{author}{\bibinfo{person}{Junnan Li}, \bibinfo{person}{Dongxu Li}, \bibinfo{person}{Silvio Savarese}, {and} \bibinfo{person}{Steven Hoi}.} \bibinfo{year}{2023}\natexlab{b}.
\newblock \showarticletitle{Blip-2: Bootstrapping language-image pre-training with frozen image encoders and large language models}. In \bibinfo{booktitle}{\emph{International Conference on Machine Learning (ICML)}}. \bibinfo{pages}{19730--19742}.
\newblock


\bibitem[Li et~al\mbox{.}(2023c)]%
        {recformer}
\bibfield{author}{\bibinfo{person}{Jiacheng Li}, \bibinfo{person}{Ming Wang}, \bibinfo{person}{Jin Li}, \bibinfo{person}{Jinmiao Fu}, \bibinfo{person}{Xin Shen}, \bibinfo{person}{Jingbo Shang}, {and} \bibinfo{person}{Julian McAuley}.} \bibinfo{year}{2023}\natexlab{c}.
\newblock \showarticletitle{Text is all you need: Learning language representations for sequential recommendation}. In \bibinfo{booktitle}{\emph{Proceedings of the 29th ACM SIGKDD Conference on Knowledge Discovery and Data Mining (KDD)}}. \bibinfo{pages}{1258--1267}.
\newblock


\bibitem[Li et~al\mbox{.}(2022)]%
        {miner}
\bibfield{author}{\bibinfo{person}{Jian Li}, \bibinfo{person}{Jieming Zhu}, \bibinfo{person}{Qiwei Bi}, \bibinfo{person}{Guohao Cai}, \bibinfo{person}{Lifeng Shang}, \bibinfo{person}{Zhenhua Dong}, \bibinfo{person}{Xin Jiang}, {and} \bibinfo{person}{Qun Liu}.} \bibinfo{year}{2022}\natexlab{}.
\newblock \showarticletitle{MINER: Multi-interest matching network for news recommendation}. In \bibinfo{booktitle}{\emph{Findings of the Association for Computational Linguistics (ACL)}}. \bibinfo{pages}{343--352}.
\newblock


\bibitem[Li et~al\mbox{.}(2023e)]%
        {li2023personalized}
\bibfield{author}{\bibinfo{person}{Lei Li}, \bibinfo{person}{Yongfeng Zhang}, {and} \bibinfo{person}{Li Chen}.} \bibinfo{year}{2023}\natexlab{e}.
\newblock \bibinfo{title}{Personalized Prompt Learning for Explainable Recommendation}.
\newblock
\newblock
\showeprint[arxiv]{2202.07371}


\bibitem[Li et~al\mbox{.}(2020)]%
        {LiWTZOOZ20}
\bibfield{author}{\bibinfo{person}{Xiang Li}, \bibinfo{person}{Chao Wang}, \bibinfo{person}{Jiwei Tan}, \bibinfo{person}{Xiaoyi Zeng}, \bibinfo{person}{Dan Ou}, {and} \bibinfo{person}{Bo Zheng}.} \bibinfo{year}{2020}\natexlab{}.
\newblock \showarticletitle{Adversarial Multimodal Representation Learning for Click-Through Rate Prediction}. In \bibinfo{booktitle}{\emph{WWW}}. \bibinfo{pages}{827--836}.
\newblock


\bibitem[Li et~al\mbox{.}(2023f)]%
        {li2023pbnr}
\bibfield{author}{\bibinfo{person}{Xinyi Li}, \bibinfo{person}{Yongfeng Zhang}, {and} \bibinfo{person}{Edward~C. Malthouse}.} \bibinfo{year}{2023}\natexlab{f}.
\newblock \showarticletitle{{PBNR:} Prompt-based News Recommender System}.
\newblock \bibinfo{journal}{\emph{CoRR}}  \bibinfo{volume}{abs/2304.07862} (\bibinfo{year}{2023}).
\newblock


\bibitem[Li and Liang(2021)]%
        {li2021prefix}
\bibfield{author}{\bibinfo{person}{Xiang~Lisa Li} {and} \bibinfo{person}{Percy Liang}.} \bibinfo{year}{2021}\natexlab{}.
\newblock \showarticletitle{Prefix-Tuning: Optimizing Continuous Prompts for Generation}. In \bibinfo{booktitle}{\emph{Proceedings of the 59th Annual Meeting of the Association for Computational Linguistics and the 11th International Joint Conference on Natural Language Processing ({ACL/IJCNLP})}}. \bibinfo{pages}{4582--4597}.
\newblock


\bibitem[Li et~al\mbox{.}(2023d)]%
        {MERT}
\bibfield{author}{\bibinfo{person}{Yizhi Li}, \bibinfo{person}{Ruibin Yuan}, \bibinfo{person}{Ge Zhang}, \bibinfo{person}{Yinghao Ma}, \bibinfo{person}{Xingran Chen}, {et~al\mbox{.}}} \bibinfo{year}{2023}\natexlab{d}.
\newblock \showarticletitle{{MERT:} Acoustic Music Understanding Model with Large-Scale Self-supervised Training}.
\newblock \bibinfo{journal}{\emph{CoRR}}  \bibinfo{volume}{abs/2306.00107} (\bibinfo{year}{2023}).
\newblock


\bibitem[Lin et~al\mbox{.}(2023a)]%
        {VideoLLaVA}
\bibfield{author}{\bibinfo{person}{Bin Lin}, \bibinfo{person}{Yang Ye}, \bibinfo{person}{Bin Zhu}, \bibinfo{person}{Jiaxi Cui}, \bibinfo{person}{Munan Ning}, \bibinfo{person}{Peng Jin}, {and} \bibinfo{person}{Li Yuan}.} \bibinfo{year}{2023}\natexlab{a}.
\newblock \showarticletitle{Video-LLaVA: Learning United Visual Representation by Alignment Before Projection}.
\newblock \bibinfo{journal}{\emph{CoRR}}  \bibinfo{volume}{abs/2311.10122} (\bibinfo{year}{2023}).
\newblock


\bibitem[Lin et~al\mbox{.}(2023b)]%
        {AutoPoster}
\bibfield{author}{\bibinfo{person}{Jinpeng Lin}, \bibinfo{person}{Min Zhou}, \bibinfo{person}{Ye Ma}, \bibinfo{person}{Yifan Gao}, \bibinfo{person}{Chenxi Fei}, \bibinfo{person}{Yangjian Chen}, \bibinfo{person}{Zhang Yu}, {and} \bibinfo{person}{Tiezheng Ge}.} \bibinfo{year}{2023}\natexlab{b}.
\newblock \showarticletitle{AutoPoster: {A} Highly Automatic and Content-aware Design System for Advertising Poster Generation}. In \bibinfo{booktitle}{\emph{ACM MM}}. \bibinfo{pages}{1250--1260}.
\newblock


\bibitem[Liu et~al\mbox{.}(2021a)]%
        {nova}
\bibfield{author}{\bibinfo{person}{Chang Liu}, \bibinfo{person}{Xiaoguang Li}, \bibinfo{person}{Guohao Cai}, \bibinfo{person}{Zhenhua Dong}, \bibinfo{person}{Hong Zhu}, {and} \bibinfo{person}{Lifeng Shang}.} \bibinfo{year}{2021}\natexlab{a}.
\newblock \showarticletitle{Noninvasive self-attention for side information fusion in sequential recommendation}. In \bibinfo{booktitle}{\emph{Proceedings of the AAAI Conference on Artificial Intelligence (AAAI)}}. \bibinfo{pages}{4249--4256}.
\newblock


\bibitem[Liu et~al\mbox{.}(2020b)]%
        {LiuYDSYDGWRXCM20}
\bibfield{author}{\bibinfo{person}{Chang Liu}, \bibinfo{person}{Han Yu}, \bibinfo{person}{Yi Dong}, \bibinfo{person}{Zhiqi Shen}, \bibinfo{person}{Yingxue Yu}, \bibinfo{person}{Ian Dixon}, \bibinfo{person}{Zhanning Gao}, \bibinfo{person}{Pan Wang}, \bibinfo{person}{Peiran Ren}, \bibinfo{person}{Xuansong Xie}, \bibinfo{person}{Lizhen Cui}, {and} \bibinfo{person}{Chunyan Miao}.} \bibinfo{year}{2020}\natexlab{b}.
\newblock \showarticletitle{Generating Engaging Promotional Videos for E-commerce Platforms (Student Abstract)}. In \bibinfo{booktitle}{\emph{AAAI}}. \bibinfo{pages}{13865--13866}.
\newblock


\bibitem[Liu et~al\mbox{.}(2023f)]%
        {liu2023recprompt}
\bibfield{author}{\bibinfo{person}{Dairui Liu}, \bibinfo{person}{Boming Yang}, \bibinfo{person}{Honghui Du}, \bibinfo{person}{Derek Greene}, \bibinfo{person}{Aonghus Lawlor}, \bibinfo{person}{Ruihai Dong}, {and} \bibinfo{person}{Irene Li}.} \bibinfo{year}{2023}\natexlab{f}.
\newblock \showarticletitle{RecPrompt: {A} Prompt Tuning Framework for News Recommendation Using Large Language Models}.
\newblock \bibinfo{journal}{\emph{CoRR}}  \bibinfo{volume}{abs/2312.10463} (\bibinfo{year}{2023}).
\newblock


\bibitem[Liu et~al\mbox{.}(2023c)]%
        {llava}
\bibfield{author}{\bibinfo{person}{Haotian Liu}, \bibinfo{person}{Chunyuan Li}, \bibinfo{person}{Qingyang Wu}, {and} \bibinfo{person}{Yong~Jae Lee}.} \bibinfo{year}{2023}\natexlab{c}.
\newblock \showarticletitle{Visual Instruction Tuning}. In \bibinfo{booktitle}{\emph{Advances in Neural Information Processing Systems (NeurIPS)}}.
\newblock


\bibitem[Liu et~al\mbox{.}(2020a)]%
        {cscnn}
\bibfield{author}{\bibinfo{person}{Hu Liu}, \bibinfo{person}{Jing Lu}, \bibinfo{person}{Hao Yang}, \bibinfo{person}{Xiwei Zhao}, \bibinfo{person}{Sulong Xu}, \bibinfo{person}{Hao Peng}, \bibinfo{person}{Zehua Zhang}, \bibinfo{person}{Wenjie Niu}, \bibinfo{person}{Xiaokun Zhu}, \bibinfo{person}{Yongjun Bao}, {et~al\mbox{.}}} \bibinfo{year}{2020}\natexlab{a}.
\newblock \showarticletitle{Category-Specific CNN for Visual-aware CTR Prediction at JD. com}. In \bibinfo{booktitle}{\emph{Proceedings of the 26th ACM SIGKDD International Conference on Knowledge Discovery \& Data Mining (KDD)}}. \bibinfo{pages}{2686--2696}.
\newblock


\bibitem[Liu et~al\mbox{.}(2023d)]%
        {mgcl}
\bibfield{author}{\bibinfo{person}{Kang Liu}, \bibinfo{person}{Feng Xue}, \bibinfo{person}{Dan Guo}, \bibinfo{person}{Peijie Sun}, \bibinfo{person}{Shengsheng Qian}, {and} \bibinfo{person}{Richang Hong}.} \bibinfo{year}{2023}\natexlab{d}.
\newblock \showarticletitle{Multimodal graph contrastive learning for multimedia-based recommendation}.
\newblock \bibinfo{journal}{\emph{IEEE Transactions on Multimedia}} (\bibinfo{year}{2023}).
\newblock


\bibitem[Liu et~al\mbox{.}(2023g)]%
        {PromptSuvey}
\bibfield{author}{\bibinfo{person}{Pengfei Liu}, \bibinfo{person}{Weizhe Yuan}, \bibinfo{person}{Jinlan Fu}, \bibinfo{person}{Zhengbao Jiang}, \bibinfo{person}{Hiroaki Hayashi}, {and} \bibinfo{person}{Graham Neubig}.} \bibinfo{year}{2023}\natexlab{g}.
\newblock \showarticletitle{Pre-train, Prompt, and Predict: {A} Systematic Survey of Prompting Methods in Natural Language Processing}.
\newblock \bibinfo{journal}{\emph{{ACM} Comput. Surv.}} \bibinfo{volume}{55}, \bibinfo{number}{9} (\bibinfo{year}{2023}), \bibinfo{pages}{195:1--195:35}.
\newblock


\bibitem[Liu et~al\mbox{.}(2024a)]%
        {once}
\bibfield{author}{\bibinfo{person}{Qijiong Liu}, \bibinfo{person}{Nuo Chen}, \bibinfo{person}{Tetsuya Sakai}, {and} \bibinfo{person}{Xiao-Ming Wu}.} \bibinfo{year}{2024}\natexlab{a}.
\newblock \showarticletitle{Once: Boosting content-based recommendation with both open-and closed-source large language models}. In \bibinfo{booktitle}{\emph{Proceedings of the 17th ACM International Conference on Web Search and Data Mining (WSDM)}}. \bibinfo{pages}{452--461}.
\newblock


\bibitem[Liu et~al\mbox{.}(2024b)]%
        {uist}
\bibfield{author}{\bibinfo{person}{Qijiong Liu}, \bibinfo{person}{Hengchang Hu}, \bibinfo{person}{Jiahao Wu}, \bibinfo{person}{Jieming Zhu}, \bibinfo{person}{Min-Yen Kan}, {and} \bibinfo{person}{Xiao-Ming Wu}.} \bibinfo{year}{2024}\natexlab{b}.
\newblock \showarticletitle{Discrete Semantic Tokenization for Deep CTR Prediction}. In \bibinfo{booktitle}{\emph{Proceedings of the ACM Web Conference (WWW)}}.
\newblock


\bibitem[Liu et~al\mbox{.}(2023a)]%
        {MMSurvey2}
\bibfield{author}{\bibinfo{person}{Qidong Liu}, \bibinfo{person}{Jiaxi Hu}, \bibinfo{person}{Yutian Xiao}, \bibinfo{person}{Jingtong Gao}, {and} \bibinfo{person}{Xiangyu Zhao}.} \bibinfo{year}{2023}\natexlab{a}.
\newblock \showarticletitle{Multimodal Recommender Systems: {A} Survey}.
\newblock \bibinfo{journal}{\emph{CoRR}}  \bibinfo{volume}{abs/2302.03883} (\bibinfo{year}{2023}).
\newblock
\urldef\tempurl%
\url{https://doi.org/10.48550/ARXIV.2302.03883}
\showDOI{\tempurl}
\showeprint[arXiv]{2302.03883}


\bibitem[Liu et~al\mbox{.}(2022b)]%
        {prec}
\bibfield{author}{\bibinfo{person}{Qijiong Liu}, \bibinfo{person}{Jieming Zhu}, \bibinfo{person}{Quanyu Dai}, {and} \bibinfo{person}{Xiao-Ming Wu}.} \bibinfo{year}{2022}\natexlab{b}.
\newblock \showarticletitle{Boosting deep CTR prediction with a plug-and-play pre-trainer for news recommendation}. In \bibinfo{booktitle}{\emph{Proceedings of the 29th International Conference on Computational Linguistics}}. \bibinfo{pages}{2823--2833}.
\newblock


\bibitem[Liu et~al\mbox{.}(2019)]%
        {liu2019user}
\bibfield{author}{\bibinfo{person}{Shang Liu}, \bibinfo{person}{Zhenzhong Chen}, \bibinfo{person}{Hongyi Liu}, {and} \bibinfo{person}{Xinghai Hu}.} \bibinfo{year}{2019}\natexlab{}.
\newblock \showarticletitle{User-video co-attention network for personalized micro-video recommendation}. In \bibinfo{booktitle}{\emph{The ACM Web Conference (WWW)}}. \bibinfo{pages}{3020--3026}.
\newblock


\bibitem[Liu et~al\mbox{.}(2023b)]%
        {Liu2023MandariMT}
\bibfield{author}{\bibinfo{person}{Xiaoqian Liu}, \bibinfo{person}{Xiuyun Li}, \bibinfo{person}{Yuan Cao}, \bibinfo{person}{Fan Zhang}, \bibinfo{person}{Xiongnan Jin}, {and} \bibinfo{person}{Jinpeng Chen}.} \bibinfo{year}{2023}\natexlab{b}.
\newblock \showarticletitle{Mandari: Multi-Modal Temporal Knowledge Graph-aware Sub-graph Embedding for Next-POI Recommendation}.
\newblock \bibinfo{journal}{\emph{IEEE International Conference on Multimedia and Expo (ICME)}} (\bibinfo{year}{2023}), \bibinfo{pages}{1529--1534}.
\newblock


\bibitem[Liu et~al\mbox{.}(2024c)]%
        {rec-gpt4v}
\bibfield{author}{\bibinfo{person}{Yuqing Liu}, \bibinfo{person}{Yu Wang}, \bibinfo{person}{Lichao Sun}, {and} \bibinfo{person}{Philip~S. Yu}.} \bibinfo{year}{2024}\natexlab{c}.
\newblock \showarticletitle{Rec-GPT4V: Multimodal Recommendation with Large Vision-Language Models}.
\newblock \bibinfo{journal}{\emph{CoRR}}  \bibinfo{volume}{abs/2402.08670} (\bibinfo{year}{2024}).
\newblock


\bibitem[Liu et~al\mbox{.}(2023e)]%
        {IDEmbedding}
\bibfield{author}{\bibinfo{person}{Yuting Liu}, \bibinfo{person}{Enneng Yang}, \bibinfo{person}{Yizhou Dang}, \bibinfo{person}{Guibing Guo}, \bibinfo{person}{Qiang Liu}, \bibinfo{person}{Yuliang Liang}, \bibinfo{person}{Linying Jiang}, {and} \bibinfo{person}{Xingwei Wang}.} \bibinfo{year}{2023}\natexlab{e}.
\newblock \showarticletitle{{ID} Embedding as Subtle Features of Content and Structure for Multimodal Recommendation}.
\newblock \bibinfo{journal}{\emph{CoRR}}  \bibinfo{volume}{abs/2311.05956} (\bibinfo{year}{2023}).
\newblock


\bibitem[Liu et~al\mbox{.}(2021b)]%
        {pmgt}
\bibfield{author}{\bibinfo{person}{Yong Liu}, \bibinfo{person}{Susen Yang}, \bibinfo{person}{Chenyi Lei}, \bibinfo{person}{Guoxin Wang}, \bibinfo{person}{Haihong Tang}, \bibinfo{person}{Juyong Zhang}, \bibinfo{person}{Aixin Sun}, {and} \bibinfo{person}{Chunyan Miao}.} \bibinfo{year}{2021}\natexlab{b}.
\newblock \showarticletitle{Pre-training graph transformer with multimodal side information for recommendation}. In \bibinfo{booktitle}{\emph{Proceedings of the 29th ACM International Conference on Multimedia (MM)}}. \bibinfo{pages}{2853--2861}.
\newblock


\bibitem[Liu et~al\mbox{.}(2024d)]%
        {Sora}
\bibfield{author}{\bibinfo{person}{Yixin Liu}, \bibinfo{person}{Kai Zhang}, \bibinfo{person}{Yuan Li}, \bibinfo{person}{Zhiling Yan}, \bibinfo{person}{Chujie Gao}, \bibinfo{person}{Ruoxi Chen}, \bibinfo{person}{Zhengqing Yuan}, \bibinfo{person}{Yue Huang}, \bibinfo{person}{Hanchi Sun}, \bibinfo{person}{Jianfeng Gao}, \bibinfo{person}{Lifang He}, {and} \bibinfo{person}{Lichao Sun}.} \bibinfo{year}{2024}\natexlab{d}.
\newblock \bibinfo{title}{Sora: A Review on Background, Technology, Limitations, and Opportunities of Large Vision Models}.
\newblock
\newblock
\showeprint[arxiv]{2402.17177}


\bibitem[Liu et~al\mbox{.}(2024e)]%
        {alignrec}
\bibfield{author}{\bibinfo{person}{Yifan Liu}, \bibinfo{person}{Kangning Zhang}, \bibinfo{person}{Xiangyuan Ren}, \bibinfo{person}{Yanhua Huang}, \bibinfo{person}{Jiarui Jin}, \bibinfo{person}{Yingjie Qin}, \bibinfo{person}{Ruilong Su}, \bibinfo{person}{Ruiwen Xu}, {and} \bibinfo{person}{Weinan Zhang}.} \bibinfo{year}{2024}\natexlab{e}.
\newblock \showarticletitle{An Aligning and Training Framework for Multimodal Recommendations}.
\newblock \bibinfo{journal}{\emph{CoRR}}  \bibinfo{volume}{abs/2403.12384} (\bibinfo{year}{2024}).
\newblock


\bibitem[Liu et~al\mbox{.}(2022a)]%
        {liu2022multi}
\bibfield{author}{\bibinfo{person}{Zhuang Liu}, \bibinfo{person}{Yunpu Ma}, \bibinfo{person}{Matthias Schubert}, \bibinfo{person}{Yuanxin Ouyang}, {and} \bibinfo{person}{Zhang Xiong}.} \bibinfo{year}{2022}\natexlab{a}.
\newblock \showarticletitle{Multi-modal contrastive pre-training for recommendation}. In \bibinfo{booktitle}{\emph{Proceedings of the 2022 International Conference on Multimedia Retrieval}}. \bibinfo{pages}{99--108}.
\newblock


\bibitem[Lu et~al\mbox{.}(2023)]%
        {Unified-IO2}
\bibfield{author}{\bibinfo{person}{Jiasen Lu}, \bibinfo{person}{Christopher Clark}, \bibinfo{person}{Sangho Lee}, \bibinfo{person}{Zichen Zhang}, \bibinfo{person}{Savya Khosla}, \bibinfo{person}{Ryan Marten}, \bibinfo{person}{Derek Hoiem}, {and} \bibinfo{person}{Aniruddha Kembhavi}.} \bibinfo{year}{2023}\natexlab{}.
\newblock \showarticletitle{Unified-IO 2: Scaling Autoregressive Multimodal Models with Vision, Language, Audio, and Action}.
\newblock \bibinfo{journal}{\emph{CoRR}}  \bibinfo{volume}{abs/2312.17172} (\bibinfo{year}{2023}).
\newblock


\bibitem[Malitesta et~al\mbox{.}(2023)]%
        {MMSurvey3}
\bibfield{author}{\bibinfo{person}{Daniele Malitesta}, \bibinfo{person}{Giandomenico Cornacchia}, \bibinfo{person}{Claudio Pomo}, \bibinfo{person}{Felice~Antonio Merra}, \bibinfo{person}{Tommaso~Di Noia}, {and} \bibinfo{person}{Eugenio~Di Sciascio}.} \bibinfo{year}{2023}\natexlab{}.
\newblock \showarticletitle{Formalizing Multimedia Recommendation through Multimodal Deep Learning}.
\newblock \bibinfo{journal}{\emph{CoRR}}  \bibinfo{volume}{abs/2309.05273} (\bibinfo{year}{2023}).
\newblock


\bibitem[Mita et~al\mbox{.}(2023)]%
        {CAMERA}
\bibfield{author}{\bibinfo{person}{Masato Mita}, \bibinfo{person}{Soichiro Murakami}, \bibinfo{person}{Akihiko Kato}, {and} \bibinfo{person}{Peinan Zhang}.} \bibinfo{year}{2023}\natexlab{}.
\newblock \showarticletitle{{CAMERA:} {A} Multimodal Dataset and Benchmark for Ad Text Generation}.
\newblock \bibinfo{journal}{\emph{CoRR}}  \bibinfo{volume}{abs/2309.12030} (\bibinfo{year}{2023}).
\newblock


\bibitem[Murakami et~al\mbox{.}(2023)]%
        {murakami2023natural}
\bibfield{author}{\bibinfo{person}{Soichiro Murakami}, \bibinfo{person}{Sho Hoshino}, {and} \bibinfo{person}{Peinan Zhang}.} \bibinfo{year}{2023}\natexlab{}.
\newblock \bibinfo{title}{Natural Language Generation for Advertising: A Survey}.
\newblock
\newblock
\showeprint[arxiv]{2306.12719}


\bibitem[Ni et~al\mbox{.}(2023)]%
        {MicroLens}
\bibfield{author}{\bibinfo{person}{Yongxin Ni}, \bibinfo{person}{Yu Cheng}, \bibinfo{person}{Xiangyan Liu}, \bibinfo{person}{Junchen Fu}, \bibinfo{person}{Youhua Li}, \bibinfo{person}{Xiangnan He}, \bibinfo{person}{Yongfeng Zhang}, {and} \bibinfo{person}{Fajie Yuan}.} \bibinfo{year}{2023}\natexlab{}.
\newblock \showarticletitle{A Content-Driven Micro-Video Recommendation Dataset at Scale}.
\newblock \bibinfo{journal}{\emph{CoRR}}  \bibinfo{volume}{abs/2309.15379} (\bibinfo{year}{2023}).
\newblock


\bibitem[OpenAI(2023a)]%
        {chatgpt}
\bibfield{author}{\bibinfo{person}{OpenAI}.} \bibinfo{year}{2023}\natexlab{a}.
\newblock \bibinfo{title}{ChatGPT}.
\newblock \bibinfo{howpublished}{\url{https://chat.openai.com/chat}}.
\newblock


\bibitem[OpenAI(2023b)]%
        {gpt-4}
\bibfield{author}{\bibinfo{person}{R OpenAI}.} \bibinfo{year}{2023}\natexlab{b}.
\newblock \showarticletitle{Gpt-4 technical report. arxiv 2303.08774}.
\newblock \bibinfo{journal}{\emph{View in Article}} \bibinfo{volume}{2}, \bibinfo{number}{5} (\bibinfo{year}{2023}).
\newblock


\bibitem[Oquab et~al\mbox{.}(2023)]%
        {DINOv2}
\bibfield{author}{\bibinfo{person}{Maxime Oquab}, \bibinfo{person}{Timoth{\'{e}}e Darcet}, \bibinfo{person}{Th{\'{e}}o Moutakanni}, \bibinfo{person}{Huy Vo}, \bibinfo{person}{Marc Szafraniec}, {et~al\mbox{.}}} \bibinfo{year}{2023}\natexlab{}.
\newblock \showarticletitle{DINOv2: Learning Robust Visual Features without Supervision}.
\newblock \bibinfo{journal}{\emph{CoRR}}  \bibinfo{volume}{abs/2304.07193} (\bibinfo{year}{2023}).
\newblock


\bibitem[Qin et~al\mbox{.}(2022)]%
        {Qin2022NextPR}
\bibfield{author}{\bibinfo{person}{Yanjun Qin}, \bibinfo{person}{Yuchen Fang}, \bibinfo{person}{Haiyong Luo}, \bibinfo{person}{Fang Zhao}, {and} \bibinfo{person}{Chenxing Wang}.} \bibinfo{year}{2022}\natexlab{}.
\newblock \showarticletitle{Next Point-of-Interest Recommendation with Auto-Correlation Enhanced Multi-Modal Transformer Network}.
\newblock \bibinfo{journal}{\emph{Proceedings of the 45th International ACM SIGIR Conference on Research and Development in Information Retrieval (SIGIR)}}.
\newblock


\bibitem[Radford et~al\mbox{.}(2021)]%
        {clip}
\bibfield{author}{\bibinfo{person}{Alec Radford}, \bibinfo{person}{Jong~Wook Kim}, \bibinfo{person}{Chris Hallacy}, \bibinfo{person}{Aditya Ramesh}, \bibinfo{person}{Gabriel Goh}, \bibinfo{person}{Sandhini Agarwal}, \bibinfo{person}{Girish Sastry}, \bibinfo{person}{Amanda Askell}, \bibinfo{person}{Pamela Mishkin}, \bibinfo{person}{Jack Clark}, {et~al\mbox{.}}} \bibinfo{year}{2021}\natexlab{}.
\newblock \showarticletitle{Learning transferable visual models from natural language supervision}. In \bibinfo{booktitle}{\emph{International Conference on Machine Learning (ICML)}}. PMLR, \bibinfo{pages}{8748--8763}.
\newblock


\bibitem[Radford et~al\mbox{.}(2018)]%
        {gpt}
\bibfield{author}{\bibinfo{person}{Alec Radford}, \bibinfo{person}{Karthik Narasimhan}, \bibinfo{person}{Tim Salimans}, \bibinfo{person}{Ilya Sutskever}, {et~al\mbox{.}}} \bibinfo{year}{2018}\natexlab{}.
\newblock \showarticletitle{Improving language understanding by generative pre-training}.
\newblock  (\bibinfo{year}{2018}).
\newblock


\bibitem[Radford et~al\mbox{.}(2019)]%
        {gpt-2}
\bibfield{author}{\bibinfo{person}{Alec Radford}, \bibinfo{person}{Jeffrey Wu}, \bibinfo{person}{Rewon Child}, \bibinfo{person}{David Luan}, \bibinfo{person}{Dario Amodei}, \bibinfo{person}{Ilya Sutskever}, {et~al\mbox{.}}} \bibinfo{year}{2019}\natexlab{}.
\newblock \showarticletitle{Language models are unsupervised multitask learners}.
\newblock \bibinfo{journal}{\emph{OpenAI blog}} \bibinfo{volume}{1}, \bibinfo{number}{8} (\bibinfo{year}{2019}), \bibinfo{pages}{9}.
\newblock


\bibitem[Raffel et~al\mbox{.}(2020)]%
        {t5}
\bibfield{author}{\bibinfo{person}{Colin Raffel}, \bibinfo{person}{Noam Shazeer}, \bibinfo{person}{Adam Roberts}, \bibinfo{person}{Katherine Lee}, \bibinfo{person}{Sharan Narang}, \bibinfo{person}{Michael Matena}, \bibinfo{person}{Yanqi Zhou}, \bibinfo{person}{Wei Li}, {and} \bibinfo{person}{Peter~J Liu}.} \bibinfo{year}{2020}\natexlab{}.
\newblock \showarticletitle{Exploring the limits of transfer learning with a unified text-to-text transformer}.
\newblock \bibinfo{journal}{\emph{Journal of Machine Learning Research}} \bibinfo{volume}{21}, \bibinfo{number}{140} (\bibinfo{year}{2020}), \bibinfo{pages}{1--67}.
\newblock


\bibitem[Rajput et~al\mbox{.}(2024)]%
        {tiger}
\bibfield{author}{\bibinfo{person}{Shashank Rajput}, \bibinfo{person}{Nikhil Mehta}, \bibinfo{person}{Anima Singh}, \bibinfo{person}{Raghunandan Hulikal~Keshavan}, \bibinfo{person}{Trung Vu}, \bibinfo{person}{Lukasz Heldt}, \bibinfo{person}{Lichan Hong}, \bibinfo{person}{Yi Tay}, \bibinfo{person}{Vinh Tran}, \bibinfo{person}{Jonah Samost}, {et~al\mbox{.}}} \bibinfo{year}{2024}\natexlab{}.
\newblock \showarticletitle{Recommender systems with generative retrieval}.
\newblock \bibinfo{journal}{\emph{Advances in Neural Information Processing Systems (NeurIPS)}}  \bibinfo{volume}{36} (\bibinfo{year}{2024}).
\newblock


\bibitem[Ramesh et~al\mbox{.}(2022)]%
        {dalle-2}
\bibfield{author}{\bibinfo{person}{Aditya Ramesh}, \bibinfo{person}{Prafulla Dhariwal}, \bibinfo{person}{Alex Nichol}, \bibinfo{person}{Casey Chu}, {and} \bibinfo{person}{Mark Chen}.} \bibinfo{year}{2022}\natexlab{}.
\newblock \showarticletitle{Hierarchical text-conditional image generation with clip latents}.
\newblock \bibinfo{journal}{\emph{arXiv preprint arXiv:2204.06125}} \bibinfo{volume}{1}, \bibinfo{number}{2} (\bibinfo{year}{2022}), \bibinfo{pages}{3}.
\newblock


\bibitem[Ramesh et~al\mbox{.}(2021)]%
        {DALLE}
\bibfield{author}{\bibinfo{person}{Aditya Ramesh}, \bibinfo{person}{Mikhail Pavlov}, \bibinfo{person}{Gabriel Goh}, \bibinfo{person}{Scott Gray}, \bibinfo{person}{Chelsea Voss}, \bibinfo{person}{Alec Radford}, \bibinfo{person}{Mark Chen}, {and} \bibinfo{person}{Ilya Sutskever}.} \bibinfo{year}{2021}\natexlab{}.
\newblock \showarticletitle{Zero-Shot Text-to-Image Generation}. In \bibinfo{booktitle}{\emph{Proceedings of the 38th International Conference on Machine Learning (ICML)}}, Vol.~\bibinfo{volume}{139}. \bibinfo{pages}{8821--8831}.
\newblock


\bibitem[Rombach et~al\mbox{.}(2022)]%
        {SD}
\bibfield{author}{\bibinfo{person}{Robin Rombach}, \bibinfo{person}{Andreas Blattmann}, \bibinfo{person}{Dominik Lorenz}, \bibinfo{person}{Patrick Esser}, {and} \bibinfo{person}{Bj{\"{o}}rn Ommer}.} \bibinfo{year}{2022}\natexlab{}.
\newblock \showarticletitle{High-Resolution Image Synthesis with Latent Diffusion Models}. In \bibinfo{booktitle}{\emph{{IEEE/CVF} Conference on Computer Vision and Pattern Recognition (CVPR)}}. \bibinfo{pages}{10674--10685}.
\newblock


\bibitem[Salemi et~al\mbox{.}(2023)]%
        {LaMP}
\bibfield{author}{\bibinfo{person}{Alireza Salemi}, \bibinfo{person}{Sheshera Mysore}, \bibinfo{person}{Michael Bendersky}, {and} \bibinfo{person}{Hamed Zamani}.} \bibinfo{year}{2023}\natexlab{}.
\newblock \showarticletitle{LaMP: When Large Language Models Meet Personalization}.
\newblock \bibinfo{journal}{\emph{CoRR}} (\bibinfo{year}{2023}).
\newblock


\bibitem[Schneider et~al\mbox{.}(2019)]%
        {wav2vec}
\bibfield{author}{\bibinfo{person}{Steffen Schneider}, \bibinfo{person}{Alexei Baevski}, \bibinfo{person}{Ronan Collobert}, {and} \bibinfo{person}{Michael Auli}.} \bibinfo{year}{2019}\natexlab{}.
\newblock \showarticletitle{wav2vec: Unsupervised Pre-Training for Speech Recognition}. In \bibinfo{booktitle}{\emph{20th Annual Conference of the International Speech Communication Association (Interspeech)}}. \bibinfo{pages}{3465--3469}.
\newblock


\bibitem[Shang et~al\mbox{.}(2023)]%
        {shang2023learning}
\bibfield{author}{\bibinfo{person}{Yu Shang}, \bibinfo{person}{Chen Gao}, \bibinfo{person}{Jiansheng Chen}, \bibinfo{person}{Depeng Jin}, \bibinfo{person}{Meng Wang}, {and} \bibinfo{person}{Yong Li}.} \bibinfo{year}{2023}\natexlab{}.
\newblock \showarticletitle{Learning fine-grained user interests for micro-video recommendation}. In \bibinfo{booktitle}{\emph{Proceedings of the 46th International ACM SIGIR Conference on Research and Development in Information Retrieval (SIGIR)}}. \bibinfo{pages}{433--442}.
\newblock


\bibitem[Shen et~al\mbox{.}(2020)]%
        {shen2020enhancing}
\bibfield{author}{\bibinfo{person}{Tiancheng Shen}, \bibinfo{person}{Jia Jia}, \bibinfo{person}{Yan Li}, \bibinfo{person}{Hanjie Wang}, {and} \bibinfo{person}{Bo Chen}.} \bibinfo{year}{2020}\natexlab{}.
\newblock \showarticletitle{Enhancing music recommendation with social media content: an attentive multimodal autoencoder approach}. In \bibinfo{booktitle}{\emph{2020 International Joint Conference on Neural Networks (IJCNN)}}. IEEE, \bibinfo{pages}{1--8}.
\newblock


\bibitem[Shen et~al\mbox{.}(2024)]%
        {PMG}
\bibfield{author}{\bibinfo{person}{Xiaoteng Shen}, \bibinfo{person}{Rui Zhang}, \bibinfo{person}{Xiaoyan Zhao}, \bibinfo{person}{Jieming Zhu}, {and} \bibinfo{person}{Xi Xiao}.} \bibinfo{year}{2024}\natexlab{}.
\newblock \showarticletitle{PMG: Personalized Multimodal Generation with Large Language Models}. In \bibinfo{booktitle}{\emph{The ACM Web Conference (WWW)}}.
\newblock


\bibitem[Shin et~al\mbox{.}(2020)]%
        {shin2020autoprompt}
\bibfield{author}{\bibinfo{person}{Taylor Shin}, \bibinfo{person}{Yasaman Razeghi}, \bibinfo{person}{Robert L.~Logan IV}, \bibinfo{person}{Eric Wallace}, {and} \bibinfo{person}{Sameer Singh}.} \bibinfo{year}{2020}\natexlab{}.
\newblock \showarticletitle{AutoPrompt: Eliciting Knowledge from Language Models with Automatically Generated Prompts}. In \bibinfo{booktitle}{\emph{Proceedings of the 2020 Conference on Empirical Methods in Natural Language Processing (EMNLP)}}. \bibinfo{pages}{4222--4235}.
\newblock


\bibitem[Singh et~al\mbox{.}(2022)]%
        {FLAVA}
\bibfield{author}{\bibinfo{person}{Amanpreet Singh}, \bibinfo{person}{Ronghang Hu}, \bibinfo{person}{Vedanuj Goswami}, \bibinfo{person}{Guillaume Couairon}, \bibinfo{person}{Wojciech Galuba}, \bibinfo{person}{Marcus Rohrbach}, {and} \bibinfo{person}{Douwe Kiela}.} \bibinfo{year}{2022}\natexlab{}.
\newblock \showarticletitle{{FLAVA:} {A} Foundational Language And Vision Alignment Model}. In \bibinfo{booktitle}{\emph{{IEEE/CVF} Conference on Computer Vision and Pattern Recognition (CVPR)}}. \bibinfo{pages}{15617--15629}.
\newblock


\bibitem[Singh et~al\mbox{.}(2023)]%
        {SemanticID}
\bibfield{author}{\bibinfo{person}{Anima Singh}, \bibinfo{person}{Trung Vu}, \bibinfo{person}{Raghunandan~H. Keshavan}, \bibinfo{person}{Nikhil Mehta}, \bibinfo{person}{Xinyang Yi}, \bibinfo{person}{Lichan Hong}, \bibinfo{person}{Lukasz Heldt}, \bibinfo{person}{Li Wei}, \bibinfo{person}{Ed~H. Chi}, {and} \bibinfo{person}{Maheswaran Sathiamoorthy}.} \bibinfo{year}{2023}\natexlab{}.
\newblock \showarticletitle{Better Generalization with Semantic IDs: {A} case study in Ranking for Recommendations}.
\newblock \bibinfo{journal}{\emph{CoRR}}  \bibinfo{volume}{abs/2306.08121} (\bibinfo{year}{2023}).
\newblock


\bibitem[Song et~al\mbox{.}(2023a)]%
        {ChatGPTKeyphrase}
\bibfield{author}{\bibinfo{person}{Mingyang Song}, \bibinfo{person}{Haiyun Jiang}, \bibinfo{person}{Shuming Shi}, \bibinfo{person}{Songfang Yao}, \bibinfo{person}{Shilong Lu}, \bibinfo{person}{Yi Feng}, \bibinfo{person}{Huafeng Liu}, {and} \bibinfo{person}{Liping Jing}.} \bibinfo{year}{2023}\natexlab{a}.
\newblock \showarticletitle{Is ChatGPT {A} Good Keyphrase Generator? {A} Preliminary Study}.
\newblock \bibinfo{journal}{\emph{CoRR}}  \bibinfo{volume}{abs/2303.13001} (\bibinfo{year}{2023}).
\newblock


\bibitem[Song et~al\mbox{.}(2023b)]%
        {Song2023MMFRecME}
\bibfield{author}{\bibinfo{person}{Xuemeng Song}, \bibinfo{person}{Chun Wang}, \bibinfo{person}{Changchang Sun}, \bibinfo{person}{Shanshan Feng}, \bibinfo{person}{Min Zhou}, {and} \bibinfo{person}{Liqiang Nie}.} \bibinfo{year}{2023}\natexlab{b}.
\newblock \showarticletitle{MM-FRec: Multi-Modal Enhanced Fashion Item Recommendation}.
\newblock \bibinfo{journal}{\emph{IEEE Transactions on Knowledge and Data Engineering}}  \bibinfo{volume}{35} (\bibinfo{year}{2023}), \bibinfo{pages}{10072--10084}.
\newblock


\bibitem[Spijkervet and Burgoyne(2021)]%
        {clmr}
\bibfield{author}{\bibinfo{person}{Janne Spijkervet} {and} \bibinfo{person}{John~Ashley Burgoyne}.} \bibinfo{year}{2021}\natexlab{}.
\newblock \showarticletitle{Contrastive Learning of Musical Representations}. In \bibinfo{booktitle}{\emph{Proceedings of the 22nd International Society for Music Information Retrieval Conference ({ISMIR})}}. \bibinfo{pages}{673--681}.
\newblock


\bibitem[Su et~al\mbox{.}(2019)]%
        {vlbert}
\bibfield{author}{\bibinfo{person}{Weijie Su}, \bibinfo{person}{Xizhou Zhu}, \bibinfo{person}{Yue Cao}, \bibinfo{person}{Bin Li}, \bibinfo{person}{Lewei Lu}, \bibinfo{person}{Furu Wei}, {and} \bibinfo{person}{Jifeng Dai}.} \bibinfo{year}{2019}\natexlab{}.
\newblock \showarticletitle{Vl-bert: Pre-training of generic visual-linguistic representations}.
\newblock \bibinfo{journal}{\emph{arXiv preprint arXiv:1908.08530}} (\bibinfo{year}{2019}).
\newblock


\bibitem[Sun et~al\mbox{.}(2019)]%
        {bert4rec}
\bibfield{author}{\bibinfo{person}{Fei Sun}, \bibinfo{person}{Jun Liu}, \bibinfo{person}{Jian Wu}, \bibinfo{person}{Changhua Pei}, \bibinfo{person}{Xiao Lin}, \bibinfo{person}{Wenwu Ou}, {and} \bibinfo{person}{Peng Jiang}.} \bibinfo{year}{2019}\natexlab{}.
\newblock \showarticletitle{BERT4Rec: Sequential recommendation with bidirectional encoder representations from transformer}. In \bibinfo{booktitle}{\emph{Proceedings of the 28th ACM International Conference on Information and Knowledge Management (CIKM)}}. \bibinfo{pages}{1441--1450}.
\newblock


\bibitem[Sun et~al\mbox{.}(2023)]%
        {sun2023universal}
\bibfield{author}{\bibinfo{person}{Wenqi Sun}, \bibinfo{person}{Ruobing Xie}, \bibinfo{person}{Shuqing Bian}, \bibinfo{person}{Wayne~Xin Zhao}, {and} \bibinfo{person}{Jie Zhou}.} \bibinfo{year}{2023}\natexlab{}.
\newblock \showarticletitle{Universal Multi-modal Multi-domain Pre-trained Recommendation}.
\newblock \bibinfo{journal}{\emph{CoRR}}  \bibinfo{volume}{abs/2311.01831} (\bibinfo{year}{2023}).
\newblock


\bibitem[Tao et~al\mbox{.}(2020)]%
        {mgat}
\bibfield{author}{\bibinfo{person}{Zhulin Tao}, \bibinfo{person}{Yinwei Wei}, \bibinfo{person}{Xiang Wang}, \bibinfo{person}{Xiangnan He}, \bibinfo{person}{Xianglin Huang}, {and} \bibinfo{person}{Tat-Seng Chua}.} \bibinfo{year}{2020}\natexlab{}.
\newblock \showarticletitle{Mgat: Multimodal graph attention network for recommendation}.
\newblock \bibinfo{journal}{\emph{Information Processing \& Management}} \bibinfo{volume}{57}, \bibinfo{number}{5} (\bibinfo{year}{2020}), \bibinfo{pages}{102277}.
\newblock


\bibitem[Touvron et~al\mbox{.}(2023a)]%
        {llama}
\bibfield{author}{\bibinfo{person}{Hugo Touvron}, \bibinfo{person}{Thibaut Lavril}, \bibinfo{person}{Gautier Izacard}, \bibinfo{person}{Xavier Martinet}, \bibinfo{person}{Marie-Anne Lachaux}, \bibinfo{person}{Timoth{\'e}e Lacroix}, \bibinfo{person}{Baptiste Rozi{\`e}re}, \bibinfo{person}{Naman Goyal}, \bibinfo{person}{Eric Hambro}, \bibinfo{person}{Faisal Azhar}, {et~al\mbox{.}}} \bibinfo{year}{2023}\natexlab{a}.
\newblock \showarticletitle{Llama: Open and efficient foundation language models}.
\newblock \bibinfo{journal}{\emph{arXiv preprint arXiv:2302.13971}} (\bibinfo{year}{2023}).
\newblock


\bibitem[Touvron et~al\mbox{.}(2023b)]%
        {llama-2}
\bibfield{author}{\bibinfo{person}{Hugo Touvron}, \bibinfo{person}{Louis Martin}, \bibinfo{person}{Kevin Stone}, \bibinfo{person}{Peter Albert}, \bibinfo{person}{Amjad Almahairi}, \bibinfo{person}{Yasmine Babaei}, \bibinfo{person}{Nikolay Bashlykov}, \bibinfo{person}{Soumya Batra}, \bibinfo{person}{Prajjwal Bhargava}, \bibinfo{person}{Shruti Bhosale}, {et~al\mbox{.}}} \bibinfo{year}{2023}\natexlab{b}.
\newblock \showarticletitle{Llama 2: Open foundation and fine-tuned chat models}.
\newblock \bibinfo{journal}{\emph{arXiv preprint arXiv:2307.09288}} (\bibinfo{year}{2023}).
\newblock


\bibitem[Trisedya et~al\mbox{.}(2022)]%
        {TrisedyaQWZ22}
\bibfield{author}{\bibinfo{person}{Bayu~Distiawan Trisedya}, \bibinfo{person}{Jianzhong Qi}, \bibinfo{person}{Wei Wang}, {and} \bibinfo{person}{Rui Zhang}.} \bibinfo{year}{2022}\natexlab{}.
\newblock \showarticletitle{{GCP:} Graph Encoder With Content-Planning for Sentence Generation From Knowledge Bases}.
\newblock \bibinfo{journal}{\emph{{IEEE} Trans. Pattern Anal. Mach. Intell.}} \bibinfo{volume}{44}, \bibinfo{number}{11} (\bibinfo{year}{2022}), \bibinfo{pages}{7521--7533}.
\newblock
\urldef\tempurl%
\url{https://doi.org/10.1109/TPAMI.2021.3118703}
\showDOI{\tempurl}


\bibitem[Tuo et~al\mbox{.}(2023)]%
        {AnyText}
\bibfield{author}{\bibinfo{person}{Yuxiang Tuo}, \bibinfo{person}{Wangmeng Xiang}, \bibinfo{person}{Jun{-}Yan He}, \bibinfo{person}{Yifeng Geng}, {and} \bibinfo{person}{Xuansong Xie}.} \bibinfo{year}{2023}\natexlab{}.
\newblock \showarticletitle{AnyText: Multilingual Visual Text Generation And Editing}.
\newblock \bibinfo{journal}{\emph{CoRR}}  \bibinfo{volume}{abs/2311.03054} (\bibinfo{year}{2023}).
\newblock


\bibitem[Van Den~Oord et~al\mbox{.}(2017)]%
        {vq-vae}
\bibfield{author}{\bibinfo{person}{Aaron Van Den~Oord}, \bibinfo{person}{Oriol Vinyals}, {et~al\mbox{.}}} \bibinfo{year}{2017}\natexlab{}.
\newblock \showarticletitle{Neural discrete representation learning}.
\newblock \bibinfo{journal}{\emph{Advances in Neural Information Processing Systems (NeurIPS)}}  \bibinfo{volume}{30} (\bibinfo{year}{2017}).
\newblock


\bibitem[Wang et~al\mbox{.}(2023b)]%
        {voyager}
\bibfield{author}{\bibinfo{person}{Guanzhi Wang}, \bibinfo{person}{Yuqi Xie}, \bibinfo{person}{Yunfan Jiang}, \bibinfo{person}{Ajay Mandlekar}, \bibinfo{person}{Chaowei Xiao}, \bibinfo{person}{Yuke Zhu}, \bibinfo{person}{Linxi Fan}, {and} \bibinfo{person}{Anima Anandkumar}.} \bibinfo{year}{2023}\natexlab{b}.
\newblock \showarticletitle{Voyager: An Open-Ended Embodied Agent with Large Language Models}.
\newblock \bibinfo{journal}{\emph{CoRR}}  \bibinfo{volume}{abs/2305.16291} (\bibinfo{year}{2023}).
\newblock


\bibitem[Wang et~al\mbox{.}(2023c)]%
        {missrec}
\bibfield{author}{\bibinfo{person}{Jinpeng Wang}, \bibinfo{person}{Ziyun Zeng}, \bibinfo{person}{Yunxiao Wang}, \bibinfo{person}{Yuting Wang}, \bibinfo{person}{Xingyu Lu}, \bibinfo{person}{Tianxiang Li}, \bibinfo{person}{Jun Yuan}, \bibinfo{person}{Rui Zhang}, \bibinfo{person}{Hai-Tao Zheng}, {and} \bibinfo{person}{Shu-Tao Xia}.} \bibinfo{year}{2023}\natexlab{c}.
\newblock \showarticletitle{Missrec: Pre-training and transferring multi-modal interest-aware sequence representation for recommendation}. In \bibinfo{booktitle}{\emph{Proceedings of the 31st ACM International Conference on Multimedia (MM)}}. \bibinfo{pages}{6548--6557}.
\newblock


\bibitem[Wang et~al\mbox{.}(2024a)]%
        {MagicVideo}
\bibfield{author}{\bibinfo{person}{Weimin Wang}, \bibinfo{person}{Jiawei Liu}, \bibinfo{person}{Zhijie Lin}, \bibinfo{person}{Jiangqiao Yan}, \bibinfo{person}{Shuo Chen}, \bibinfo{person}{Chetwin Low}, \bibinfo{person}{Tuyen Hoang}, \bibinfo{person}{Jie Wu}, \bibinfo{person}{Jun~Hao Liew}, \bibinfo{person}{Hanshu Yan}, \bibinfo{person}{Daquan Zhou}, {and} \bibinfo{person}{Jiashi Feng}.} \bibinfo{year}{2024}\natexlab{a}.
\newblock \showarticletitle{MagicVideo-V2: Multi-Stage High-Aesthetic Video Generation}.
\newblock \bibinfo{journal}{\emph{CoRR}}  \bibinfo{volume}{abs/2401.04468} (\bibinfo{year}{2024}).
\newblock


\bibitem[Wang et~al\mbox{.}(2024b)]%
        {eager}
\bibfield{author}{\bibinfo{person}{Ye Wang}, \bibinfo{person}{Jiahao Xun}, \bibinfo{person}{Mingjie Hong}, \bibinfo{person}{Jieming Zhu}, \bibinfo{person}{Tao Jin}, \bibinfo{person}{Wang Lin}, \bibinfo{person}{Haoyuan Li}, \bibinfo{person}{Linjun Li}, \bibinfo{person}{Yan Xia}, \bibinfo{person}{Zhou Zhao}, {and} \bibinfo{person}{Zhenhua Dong}.} \bibinfo{year}{2024}\natexlab{b}.
\newblock \showarticletitle{EAGER: Two-Stream Generative Recommender with Behavior-Semantic Collaboration}. In \bibinfo{booktitle}{\emph{Proceedings of the ACM SIGKDD Conference on Knowledge Discovery and Data Mining (KDD)}}.
\newblock


\bibitem[Wang et~al\mbox{.}(2023a)]%
        {WangTRCWA23}
\bibfield{author}{\bibinfo{person}{Zhenduo Wang}, \bibinfo{person}{Yuancheng Tu}, \bibinfo{person}{Corby Rosset}, \bibinfo{person}{Nick Craswell}, \bibinfo{person}{Ming Wu}, {and} \bibinfo{person}{Qingyao Ai}.} \bibinfo{year}{2023}\natexlab{a}.
\newblock \showarticletitle{Zero-shot Clarifying Question Generation for Conversational Search}. In \bibinfo{booktitle}{\emph{Proceedings of the ACM Web Conference (WWW)}}. \bibinfo{pages}{3288--3298}.
\newblock


\bibitem[Wei et~al\mbox{.}(2024a)]%
        {DeepMP}
\bibfield{author}{\bibinfo{person}{Tianxin Wei}, \bibinfo{person}{Bowen Jin}, \bibinfo{person}{Ruirui Li}, \bibinfo{person}{Hansi Zeng}, \bibinfo{person}{Zhengyang Wang}, \bibinfo{person}{Jianhui Sun}, \bibinfo{person}{Qingyu Yin}, \bibinfo{person}{Hanqing Lu}, \bibinfo{person}{Suhang Wang}, \bibinfo{person}{Jingrui He}, {and} \bibinfo{person}{Xianfeng Tang}.} \bibinfo{year}{2024}\natexlab{a}.
\newblock \showarticletitle{Towards Unified Multi-Modal Personalization: Large Vision-Language Models for Generative Recommendation and Beyond}.
\newblock \bibinfo{journal}{\emph{CoRR}} (\bibinfo{year}{2024}).
\newblock


\bibitem[Wei et~al\mbox{.}(2023a)]%
        {mmssl}
\bibfield{author}{\bibinfo{person}{Wei Wei}, \bibinfo{person}{Chao Huang}, \bibinfo{person}{Lianghao Xia}, {and} \bibinfo{person}{Chuxu Zhang}.} \bibinfo{year}{2023}\natexlab{a}.
\newblock \showarticletitle{Multi-modal self-supervised learning for recommendation}. In \bibinfo{booktitle}{\emph{Proceedings of the ACM Web Conference 2023}}. \bibinfo{pages}{790--800}.
\newblock


\bibitem[Wei et~al\mbox{.}(2024b)]%
        {wei2024promptmm}
\bibfield{author}{\bibinfo{person}{Wei Wei}, \bibinfo{person}{Jiabin Tang}, \bibinfo{person}{Lianghao Xia}, \bibinfo{person}{Yangqin Jiang}, {and} \bibinfo{person}{Chao Huang}.} \bibinfo{year}{2024}\natexlab{b}.
\newblock \showarticletitle{PromptMM: Multi-Modal Knowledge Distillation for Recommendation with Prompt-Tuning}. In \bibinfo{booktitle}{\emph{Proceedings of the {ACM} on Web Conference (WWW)}}. \bibinfo{pages}{3217--3228}.
\newblock


\bibitem[Wei et~al\mbox{.}(2023b)]%
        {lightgt}
\bibfield{author}{\bibinfo{person}{Yinwei Wei}, \bibinfo{person}{Wenqi Liu}, \bibinfo{person}{Fan Liu}, \bibinfo{person}{Xiang Wang}, \bibinfo{person}{Liqiang Nie}, {and} \bibinfo{person}{Tat-Seng Chua}.} \bibinfo{year}{2023}\natexlab{b}.
\newblock \showarticletitle{Lightgt: A light graph transformer for multimedia recommendation}. In \bibinfo{booktitle}{\emph{Proceedings of the 46th International ACM SIGIR Conference on Research and Development in Information Retrieval (SIGIR)}}. \bibinfo{pages}{1508--1517}.
\newblock


\bibitem[Wei et~al\mbox{.}(2021)]%
        {CLCRec}
\bibfield{author}{\bibinfo{person}{Yinwei Wei}, \bibinfo{person}{Xiang Wang}, \bibinfo{person}{Qi Li}, \bibinfo{person}{Liqiang Nie}, \bibinfo{person}{Yan Li}, \bibinfo{person}{Xuanping Li}, {and} \bibinfo{person}{Tat{-}Seng Chua}.} \bibinfo{year}{2021}\natexlab{}.
\newblock \showarticletitle{Contrastive Learning for Cold-Start Recommendation}.
\newblock \bibinfo{journal}{\emph{CoRR}}  \bibinfo{volume}{abs/2107.05315} (\bibinfo{year}{2021}).
\newblock


\bibitem[Wei et~al\mbox{.}(2020)]%
        {grcn}
\bibfield{author}{\bibinfo{person}{Yinwei Wei}, \bibinfo{person}{Xiang Wang}, \bibinfo{person}{Liqiang Nie}, \bibinfo{person}{Xiangnan He}, {and} \bibinfo{person}{Tat{-}Seng Chua}.} \bibinfo{year}{2020}\natexlab{}.
\newblock \showarticletitle{Graph-Refined Convolutional Network for Multimedia Recommendation with Implicit Feedback}. In \bibinfo{booktitle}{\emph{The 28th {ACM} International Conference on Multimedia (MM)}}. \bibinfo{pages}{3541--3549}.
\newblock


\bibitem[Wei et~al\mbox{.}(2019)]%
        {mmgcn}
\bibfield{author}{\bibinfo{person}{Yinwei Wei}, \bibinfo{person}{Xiang Wang}, \bibinfo{person}{Liqiang Nie}, \bibinfo{person}{Xiangnan He}, \bibinfo{person}{Richang Hong}, {and} \bibinfo{person}{Tat-Seng Chua}.} \bibinfo{year}{2019}\natexlab{}.
\newblock \showarticletitle{MMGCN: Multi-modal graph convolution network for personalized recommendation of micro-video}. In \bibinfo{booktitle}{\emph{Proceedings of the 27th ACM International Conference on Multimedia (MM)}}. \bibinfo{pages}{1437--1445}.
\newblock


\bibitem[Wu et~al\mbox{.}(2019)]%
        {nrms}
\bibfield{author}{\bibinfo{person}{Chuhan Wu}, \bibinfo{person}{Fangzhao Wu}, \bibinfo{person}{Suyu Ge}, \bibinfo{person}{Tao Qi}, \bibinfo{person}{Yongfeng Huang}, {and} \bibinfo{person}{Xing Xie}.} \bibinfo{year}{2019}\natexlab{}.
\newblock \showarticletitle{Neural news recommendation with multi-head self-attention}. In \bibinfo{booktitle}{\emph{Proceedings of the Conference on Empirical Methods in Natural Language Processing and the International Joint Conference on Natural Language Processing (EMNLP-IJCNLP)}}. \bibinfo{pages}{6389--6394}.
\newblock


\bibitem[Wu et~al\mbox{.}(2021)]%
        {plm-nr}
\bibfield{author}{\bibinfo{person}{Chuhan Wu}, \bibinfo{person}{Fangzhao Wu}, \bibinfo{person}{Tao Qi}, {and} \bibinfo{person}{Yongfeng Huang}.} \bibinfo{year}{2021}\natexlab{}.
\newblock \showarticletitle{Empowering news recommendation with pre-trained language models}. In \bibinfo{booktitle}{\emph{Proceedings of the 44th international ACM SIGIR Conference on Research and Development in Information Retrieval (SIGIR)}}. \bibinfo{pages}{1652--1656}.
\newblock


\bibitem[Wu et~al\mbox{.}(2022)]%
        {mmrec}
\bibfield{author}{\bibinfo{person}{Chuhan Wu}, \bibinfo{person}{Fangzhao Wu}, \bibinfo{person}{Tao Qi}, \bibinfo{person}{Chao Zhang}, \bibinfo{person}{Yongfeng Huang}, {and} \bibinfo{person}{Tong Xu}.} \bibinfo{year}{2022}\natexlab{}.
\newblock \showarticletitle{MM-Rec: Visiolinguistic Model Empowered Multimodal News Recommendation}. In \bibinfo{booktitle}{\emph{The 45th International {ACM} {SIGIR} Conference on Research and Development in Information Retrieval (SIGIR)}}. \bibinfo{pages}{2560--2564}.
\newblock


\bibitem[Xi et~al\mbox{.}(2023)]%
        {xi2023towards}
\bibfield{author}{\bibinfo{person}{Yunjia Xi}, \bibinfo{person}{Weiwen Liu}, \bibinfo{person}{Jianghao Lin}, \bibinfo{person}{Jieming Zhu}, \bibinfo{person}{Bo Chen}, \bibinfo{person}{Ruiming Tang}, \bibinfo{person}{Weinan Zhang}, \bibinfo{person}{Rui Zhang}, {and} \bibinfo{person}{Yong Yu}.} \bibinfo{year}{2023}\natexlab{}.
\newblock \showarticletitle{Towards open-world recommendation with knowledge augmentation from large language models}.
\newblock \bibinfo{journal}{\emph{arXiv preprint arXiv:2306.10933}} (\bibinfo{year}{2023}).
\newblock


\bibitem[Xiao et~al\mbox{.}(2022a)]%
        {XiaoDCJYDL22}
\bibfield{author}{\bibinfo{person}{Fangxiong Xiao}, \bibinfo{person}{Lixi Deng}, \bibinfo{person}{Jingjing Chen}, \bibinfo{person}{Houye Ji}, \bibinfo{person}{Xiaorui Yang}, \bibinfo{person}{Zhuoye Ding}, {and} \bibinfo{person}{Bo Long}.} \bibinfo{year}{2022}\natexlab{a}.
\newblock \showarticletitle{From Abstract to Details: {A} Generative Multimodal Fusion Framework for Recommendation}. In \bibinfo{booktitle}{\emph{MM}}. \bibinfo{pages}{258--267}.
\newblock


\bibitem[Xiao et~al\mbox{.}(2022b)]%
        {SpeedyFeed}
\bibfield{author}{\bibinfo{person}{Shitao Xiao}, \bibinfo{person}{Zheng Liu}, \bibinfo{person}{Yingxia Shao}, \bibinfo{person}{Tao Di}, \bibinfo{person}{Bhuvan Middha}, \bibinfo{person}{Fangzhao Wu}, {and} \bibinfo{person}{Xing Xie}.} \bibinfo{year}{2022}\natexlab{b}.
\newblock \showarticletitle{Training Large-Scale News Recommenders with Pretrained Language Models in the Loop}. In \bibinfo{booktitle}{\emph{The 28th {ACM} {SIGKDD} Conference on Knowledge Discovery and Data Mining (KDD)}}. \bibinfo{pages}{4215--4225}.
\newblock


\bibitem[Xu et~al\mbox{.}(2024)]%
        {xu2024prompting}
\bibfield{author}{\bibinfo{person}{Lanling Xu}, \bibinfo{person}{Junjie Zhang}, \bibinfo{person}{Bingqian Li}, \bibinfo{person}{Jinpeng Wang}, \bibinfo{person}{Mingchen Cai}, \bibinfo{person}{Wayne~Xin Zhao}, {and} \bibinfo{person}{Ji{-}Rong Wen}.} \bibinfo{year}{2024}\natexlab{}.
\newblock \showarticletitle{Prompting Large Language Models for Recommender Systems: {A} Comprehensive Framework and Empirical Analysis}.
\newblock \bibinfo{journal}{\emph{CoRR}}  \bibinfo{volume}{abs/2401.04997} (\bibinfo{year}{2024}).
\newblock


\bibitem[Xu et~al\mbox{.}(2021)]%
        {KPLUG}
\bibfield{author}{\bibinfo{person}{Song Xu}, \bibinfo{person}{Haoran Li}, \bibinfo{person}{Peng Yuan}, \bibinfo{person}{Yujia Wang}, \bibinfo{person}{Youzheng Wu}, \bibinfo{person}{Xiaodong He}, \bibinfo{person}{Ying Liu}, {and} \bibinfo{person}{Bowen Zhou}.} \bibinfo{year}{2021}\natexlab{}.
\newblock \showarticletitle{{K-PLUG:} Knowledge-injected Pre-trained Language Model for Natural Language Understanding and Generation in E-Commerce}. In \bibinfo{booktitle}{\emph{Findings of EMNLP}}. \bibinfo{pages}{1--17}.
\newblock


\bibitem[Xun et~al\mbox{.}(2021)]%
        {im-rec}
\bibfield{author}{\bibinfo{person}{Jiahao Xun}, \bibinfo{person}{Shengyu Zhang}, \bibinfo{person}{Zhou Zhao}, \bibinfo{person}{Jieming Zhu}, \bibinfo{person}{Qi Zhang}, \bibinfo{person}{Jingjie Li}, \bibinfo{person}{Xiuqiang He}, \bibinfo{person}{Xiaofei He}, \bibinfo{person}{Tat-Seng Chua}, {and} \bibinfo{person}{Fei Wu}.} \bibinfo{year}{2021}\natexlab{}.
\newblock \showarticletitle{Why do we click: visual impression-aware news recommendation}. In \bibinfo{booktitle}{\emph{Proceedings of the 29th ACM International Conference on Multimedia (MM)}}. \bibinfo{pages}{3881--3890}.
\newblock


\bibitem[Xv et~al\mbox{.}(2022)]%
        {xv2022visual}
\bibfield{author}{\bibinfo{person}{Guipeng Xv}, \bibinfo{person}{Si Chen}, \bibinfo{person}{Chen Lin}, \bibinfo{person}{Wanxian Guan}, \bibinfo{person}{Xingyuan Bu}, \bibinfo{person}{Xubin Li}, \bibinfo{person}{Hongbo Deng}, \bibinfo{person}{Jian Xu}, {and} \bibinfo{person}{Bo Zheng}.} \bibinfo{year}{2022}\natexlab{}.
\newblock \showarticletitle{Visual Encoding and Debiasing for CTR Prediction}. In \bibinfo{booktitle}{\emph{Proceedings of the 31st ACM International Conference on Information \& Knowledge Management (CIKM)}}. \bibinfo{pages}{4615--4619}.
\newblock


\bibitem[Yang et~al\mbox{.}(2022)]%
        {YangZEL22}
\bibfield{author}{\bibinfo{person}{Shiquan Yang}, \bibinfo{person}{Rui Zhang}, \bibinfo{person}{Sarah~M. Erfani}, {and} \bibinfo{person}{Jey~Han Lau}.} \bibinfo{year}{2022}\natexlab{}.
\newblock \showarticletitle{An Interpretable Neuro-Symbolic Reasoning Framework for Task-Oriented Dialogue Generation}. In \bibinfo{booktitle}{\emph{Proceedings of the 60th Annual Meeting of the Association for Computational Linguistics (ACL)}}. \bibinfo{pages}{4918--4935}.
\newblock


\bibitem[Yang et~al\mbox{.}(2019)]%
        {YangDTTZQD19}
\bibfield{author}{\bibinfo{person}{Xiao Yang}, \bibinfo{person}{Tao Deng}, \bibinfo{person}{Weihan Tan}, \bibinfo{person}{Xutian Tao}, \bibinfo{person}{Junwei Zhang}, \bibinfo{person}{Shouke Qin}, {and} \bibinfo{person}{Zongyao Ding}.} \bibinfo{year}{2019}\natexlab{}.
\newblock \showarticletitle{Learning Compositional, Visual and Relational Representations for {CTR} Prediction in Sponsored Search}. In \bibinfo{booktitle}{\emph{CIKM}}. \bibinfo{pages}{2851--2859}.
\newblock


\bibitem[Yang et~al\mbox{.}(2023)]%
        {ParallelRanking}
\bibfield{author}{\bibinfo{person}{Zhiguang Yang}, \bibinfo{person}{Lu Wang}, \bibinfo{person}{Chun Gan}, \bibinfo{person}{Liufang Sang}, {and} \bibinfo{person}{et al.}} \bibinfo{year}{2023}\natexlab{}.
\newblock \showarticletitle{Parallel Ranking of Ads and Creatives in Real-Time Advertising Systems}.
\newblock \bibinfo{journal}{\emph{CoRR}} (\bibinfo{year}{2023}).
\newblock


\bibitem[Yao et~al\mbox{.}(2024)]%
        {MART}
\bibfield{author}{\bibinfo{person}{Dong Yao}, \bibinfo{person}{Jieming Zhu}, \bibinfo{person}{Jiahao Xun}, \bibinfo{person}{Shengyu Zhang}, \bibinfo{person}{Zhou Zhao}, \bibinfo{person}{Liqun Deng}, \bibinfo{person}{Wenqiao Zhang}, \bibinfo{person}{Zhenhua Dong}, {and} \bibinfo{person}{Xin Jiang}.} \bibinfo{year}{2024}\natexlab{}.
\newblock \showarticletitle{{MART:} Learning Hierarchical Music Audio Representations with Part-Whole Transformer}. In \bibinfo{booktitle}{\emph{Companion Proceedings of the {ACM} on Web Conference (WWW)}}. \bibinfo{pages}{967--970}.
\newblock


\bibitem[Yi et~al\mbox{.}(2022)]%
        {multimodalGC}
\bibfield{author}{\bibinfo{person}{Zixuan Yi}, \bibinfo{person}{Xi Wang}, \bibinfo{person}{Iadh Ounis}, {and} \bibinfo{person}{Craig Macdonald}.} \bibinfo{year}{2022}\natexlab{}.
\newblock \showarticletitle{Multi-modal Graph Contrastive Learning for Micro-video Recommendation}.
\newblock \bibinfo{journal}{\emph{Proceedings of the 45th International ACM SIGIR Conference on Research and Development in Information Retrieval (SIGIR)}} (\bibinfo{year}{2022}).
\newblock


\bibitem[Yu et~al\mbox{.}(2022b)]%
        {coca}
\bibfield{author}{\bibinfo{person}{Jiahui Yu}, \bibinfo{person}{Zirui Wang}, \bibinfo{person}{Vijay Vasudevan}, \bibinfo{person}{Legg Yeung}, \bibinfo{person}{Mojtaba Seyedhosseini}, {and} \bibinfo{person}{Yonghui Wu}.} \bibinfo{year}{2022}\natexlab{b}.
\newblock \showarticletitle{CoCa: Contrastive Captioners are Image-Text Foundation Models}.
\newblock \bibinfo{journal}{\emph{Trans. Mach. Learn. Res.}}  \bibinfo{volume}{2022} (\bibinfo{year}{2022}).
\newblock


\bibitem[Yu et~al\mbox{.}(2022a)]%
        {CommerceMM}
\bibfield{author}{\bibinfo{person}{Licheng Yu}, \bibinfo{person}{Jun Chen}, \bibinfo{person}{Animesh Sinha}, \bibinfo{person}{Mengjiao Wang}, \bibinfo{person}{Yu Chen}, \bibinfo{person}{Tamara~L. Berg}, {and} \bibinfo{person}{Ning Zhang}.} \bibinfo{year}{2022}\natexlab{a}.
\newblock \showarticletitle{CommerceMM: Large-Scale Commerce MultiModal Representation Learning with Omni Retrieval}. In \bibinfo{booktitle}{\emph{The 28th {ACM} {SIGKDD} Conference on Knowledge Discovery and Data Mining (KDD)}}. \bibinfo{pages}{4433--4442}.
\newblock


\bibitem[Yuan et~al\mbox{.}(2023)]%
        {YuanYSLFYPN23}
\bibfield{author}{\bibinfo{person}{Zheng Yuan}, \bibinfo{person}{Fajie Yuan}, \bibinfo{person}{Yu Song}, \bibinfo{person}{Youhua Li}, \bibinfo{person}{Junchen Fu}, \bibinfo{person}{Fei Yang}, \bibinfo{person}{Yunzhu Pan}, {and} \bibinfo{person}{Yongxin Ni}.} \bibinfo{year}{2023}\natexlab{}.
\newblock \showarticletitle{Where to go next for recommender systems? id-vs. modality-based recommender models revisited}. In \bibinfo{booktitle}{\emph{Proceedings of the 46th International ACM SIGIR Conference on Research and Development in Information Retrieval (SIGIR)}}. \bibinfo{pages}{2639--2649}.
\newblock


\bibitem[Zeng et~al\mbox{.}(2021)]%
        {musicbert}
\bibfield{author}{\bibinfo{person}{Mingliang Zeng}, \bibinfo{person}{Xu Tan}, \bibinfo{person}{Rui Wang}, \bibinfo{person}{Zeqian Ju}, \bibinfo{person}{Tao Qin}, {and} \bibinfo{person}{Tie-Yan Liu}.} \bibinfo{year}{2021}\natexlab{}.
\newblock \showarticletitle{MusicBERT: Symbolic Music Understanding with Large-Scale Pre-Training}. In \bibinfo{booktitle}{\emph{Findings of the Association for Computational Linguistics (ACL)}}. \bibinfo{pages}{791--800}.
\newblock


\bibitem[Zhan et~al\mbox{.}(2024)]%
        {AnyGPT}
\bibfield{author}{\bibinfo{person}{Jun Zhan}, \bibinfo{person}{Junqi Dai}, \bibinfo{person}{Jiasheng Ye}, \bibinfo{person}{Yunhua Zhou}, {et~al\mbox{.}}} \bibinfo{year}{2024}\natexlab{}.
\newblock \showarticletitle{AnyGPT: Unified Multimodal {LLM} with Discrete Sequence Modeling}.
\newblock \bibinfo{journal}{\emph{CoRR}}  \bibinfo{volume}{abs/2402.12226} (\bibinfo{year}{2024}).
\newblock


\bibitem[Zhang et~al\mbox{.}(2023b)]%
        {msm4sr}
\bibfield{author}{\bibinfo{person}{Lingzi Zhang}, \bibinfo{person}{Xin Zhou}, {and} \bibinfo{person}{Zhiqi Shen}.} \bibinfo{year}{2023}\natexlab{b}.
\newblock \showarticletitle{Multimodal pre-training framework for sequential recommendation via contrastive learning}.
\newblock \bibinfo{journal}{\emph{arXiv preprint arXiv:2303.11879}} (\bibinfo{year}{2023}).
\newblock


\bibitem[Zhang et~al\mbox{.}(2021)]%
        {unbert}
\bibfield{author}{\bibinfo{person}{Qi Zhang}, \bibinfo{person}{Jingjie Li}, \bibinfo{person}{Qinglin Jia}, \bibinfo{person}{Chuyuan Wang}, \bibinfo{person}{Jieming Zhu}, \bibinfo{person}{Zhaowei Wang}, {and} \bibinfo{person}{Xiuqiang He}.} \bibinfo{year}{2021}\natexlab{}.
\newblock \showarticletitle{UNBERT: User-News Matching BERT for News Recommendation.}. In \bibinfo{booktitle}{\emph{IJCAI}}, Vol.~\bibinfo{volume}{21}. \bibinfo{pages}{3356--3362}.
\newblock


\bibitem[Zhang et~al\mbox{.}(2023a)]%
        {Meta-Transformer}
\bibfield{author}{\bibinfo{person}{Yiyuan Zhang}, \bibinfo{person}{Kaixiong Gong}, \bibinfo{person}{Kaipeng Zhang}, \bibinfo{person}{Hongsheng Li}, \bibinfo{person}{Yu Qiao}, \bibinfo{person}{Wanli Ouyang}, {and} \bibinfo{person}{Xiangyu Yue}.} \bibinfo{year}{2023}\natexlab{a}.
\newblock \showarticletitle{Meta-Transformer: {A} Unified Framework for Multimodal Learning}.
\newblock \bibinfo{journal}{\emph{CoRR}}  \bibinfo{volume}{abs/2307.10802} (\bibinfo{year}{2023}).
\newblock


\bibitem[Zhang and Wang(2023)]%
        {Prompt4NR}
\bibfield{author}{\bibinfo{person}{Zizhuo Zhang} {and} \bibinfo{person}{Bang Wang}.} \bibinfo{year}{2023}\natexlab{}.
\newblock \showarticletitle{Prompt Learning for News Recommendation}. In \bibinfo{booktitle}{\emph{Proceedings of the 46th International {ACM} {SIGIR} Conference on Research and Development in Information Retrieval (SIGIR)}}. \bibinfo{pages}{227--237}.
\newblock


\bibitem[Zhao et~al\mbox{.}(2019)]%
        {ZhaoFSSCXQ19}
\bibfield{author}{\bibinfo{person}{Guoshuai Zhao}, \bibinfo{person}{Hao Fu}, \bibinfo{person}{Ruihua Song}, \bibinfo{person}{Tetsuya Sakai}, \bibinfo{person}{Zhongxia Chen}, \bibinfo{person}{Xing Xie}, {and} \bibinfo{person}{Xueming Qian}.} \bibinfo{year}{2019}\natexlab{}.
\newblock \showarticletitle{Personalized Reason Generation for Explainable Song Recommendation}.
\newblock \bibinfo{journal}{\emph{{ACM} Trans. Intell. Syst. Technol.}} \bibinfo{volume}{10}, \bibinfo{number}{4} (\bibinfo{year}{2019}), \bibinfo{pages}{41:1--41:21}.
\newblock


\bibitem[Zhou et~al\mbox{.}(2023b)]%
        {MMSurvey}
\bibfield{author}{\bibinfo{person}{Hongyu Zhou}, \bibinfo{person}{Xin Zhou}, \bibinfo{person}{Zhiwei Zeng}, \bibinfo{person}{Lingzi Zhang}, {and} \bibinfo{person}{Zhiqi Shen}.} \bibinfo{year}{2023}\natexlab{b}.
\newblock \showarticletitle{A Comprehensive Survey on Multimodal Recommender Systems: Taxonomy, Evaluation, and Future Directions}.
\newblock \bibinfo{journal}{\emph{CoRR}}  \bibinfo{volume}{abs/2302.04473} (\bibinfo{year}{2023}).
\newblock


\bibitem[Zhou et~al\mbox{.}(2024)]%
        {zhou2024gcof}
\bibfield{author}{\bibinfo{person}{Jianghui Zhou}, \bibinfo{person}{Ya Gao}, \bibinfo{person}{Jie Liu}, \bibinfo{person}{Xuemin Zhao}, \bibinfo{person}{Zhaohua Yang}, \bibinfo{person}{Yue Wu}, {and} \bibinfo{person}{Lirong Shi}.} \bibinfo{year}{2024}\natexlab{}.
\newblock \bibinfo{title}{GCOF: Self-iterative Text Generation for Copywriting Using Large Language Model}.
\newblock
\newblock
\showeprint{2402.13667}


\bibitem[Zhou et~al\mbox{.}(2022)]%
        {zhou2022learning}
\bibfield{author}{\bibinfo{person}{Kaiyang Zhou}, \bibinfo{person}{Jingkang Yang}, \bibinfo{person}{Chen~Change Loy}, {and} \bibinfo{person}{Ziwei Liu}.} \bibinfo{year}{2022}\natexlab{}.
\newblock \showarticletitle{Learning to Prompt for Vision-Language Models}.
\newblock \bibinfo{journal}{\emph{Int. J. Comput. Vis.}} \bibinfo{volume}{130}, \bibinfo{number}{9} (\bibinfo{year}{2022}), \bibinfo{pages}{2337--2348}.
\newblock


\bibitem[Zhou et~al\mbox{.}(2023a)]%
        {ZhouJWCS23}
\bibfield{author}{\bibinfo{person}{Wangchunshu Zhou}, \bibinfo{person}{Yuchen~Eleanor Jiang}, \bibinfo{person}{Ethan Wilcox}, \bibinfo{person}{Ryan Cotterell}, {and} \bibinfo{person}{Mrinmaya Sachan}.} \bibinfo{year}{2023}\natexlab{a}.
\newblock \showarticletitle{Controlled Text Generation with Natural Language Instructions}. In \bibinfo{booktitle}{\emph{Proceedings of International Conference on Machine Learning (ICML)}}. \bibinfo{pages}{42602--42613}.
\newblock


\bibitem[Zhou and Shen(2023)]%
        {freedom}
\bibfield{author}{\bibinfo{person}{Xin Zhou} {and} \bibinfo{person}{Zhiqi Shen}.} \bibinfo{year}{2023}\natexlab{}.
\newblock \showarticletitle{A tale of two graphs: Freezing and denoising graph structures for multimodal recommendation}. In \bibinfo{booktitle}{\emph{Proceedings of the 31st ACM International Conference on Multimedia (MM)}}. \bibinfo{pages}{935--943}.
\newblock


\bibitem[Zhu et~al\mbox{.}(2022)]%
        {BARS}
\bibfield{author}{\bibinfo{person}{Jieming Zhu}, \bibinfo{person}{Quanyu Dai}, \bibinfo{person}{Liangcai Su}, \bibinfo{person}{Rong Ma}, \bibinfo{person}{Jinyang Liu}, \bibinfo{person}{Guohao Cai}, \bibinfo{person}{Xi Xiao}, {and} \bibinfo{person}{Rui Zhang}.} \bibinfo{year}{2022}\natexlab{}.
\newblock \showarticletitle{{BARS:} Towards Open Benchmarking for Recommender Systems}. In \bibinfo{booktitle}{\emph{The 45th International {ACM} {SIGIR} Conference on Research and Development in Information Retrieval (SIGIR)}}. \bibinfo{pages}{2912--2923}.
\newblock


\bibitem[Zhu et~al\mbox{.}(2024)]%
        {mmrec_tutorial}
\bibfield{author}{\bibinfo{person}{Jieming Zhu}, \bibinfo{person}{Xin Zhou}, \bibinfo{person}{Chuhan Wu}, \bibinfo{person}{Rui Zhang}, {and} \bibinfo{person}{Zhenhua Dong}.} \bibinfo{year}{2024}\natexlab{}.
\newblock \showarticletitle{Multimodal Pretraining and Generation for Recommendation: {A} Tutorial}. In \bibinfo{booktitle}{\emph{Companion Proceedings of the {ACM} on Web Conference 2024 (WWW)}}. \bibinfo{pages}{1272--1275}.
\newblock


\bibitem[Zhu et~al\mbox{.}(2023)]%
        {ZhuYZRCS0K23}
\bibfield{author}{\bibinfo{person}{Luyang Zhu}, \bibinfo{person}{Dawei Yang}, \bibinfo{person}{Tyler Zhu}, \bibinfo{person}{Fitsum Reda}, \bibinfo{person}{William Chan}, \bibinfo{person}{Chitwan Saharia}, \bibinfo{person}{Mohammad Norouzi}, {and} \bibinfo{person}{Ira Kemelmacher{-}Shlizerman}.} \bibinfo{year}{2023}\natexlab{}.
\newblock \showarticletitle{TryOnDiffusion: {A} Tale of Two UNets}. In \bibinfo{booktitle}{\emph{{IEEE/CVF} Conference on Computer Vision and Pattern Recognition (CVPR)}}. \bibinfo{pages}{4606--4615}.
\newblock


\bibitem[Zhu et~al\mbox{.}(2021)]%
        {K3M}
\bibfield{author}{\bibinfo{person}{Yushan Zhu}, \bibinfo{person}{Huaixiao Zhao}, \bibinfo{person}{Wen Zhang}, \bibinfo{person}{Ganqiang Ye}, \bibinfo{person}{Hui Chen}, \bibinfo{person}{Ningyu Zhang}, {and} \bibinfo{person}{Huajun Chen}.} \bibinfo{year}{2021}\natexlab{}.
\newblock \showarticletitle{Knowledge Perceived Multi-modal Pretraining in E-commerce}. In \bibinfo{booktitle}{\emph{{ACM} Multimedia Conference (MM)}}. \bibinfo{pages}{2744--2752}.
\newblock


\end{thebibliography}
\end{document}